\documentclass[sigplan,nonacm]{acmart}

\newcommand{\asplossubmissionnumber}{443}

\usepackage[normalem]{ulem}
\usepackage{xspace}
\usepackage{listings}
\usepackage{color}
\usepackage{flushend}
\usepackage{hyperref}
\usepackage{enumitem}
\usepackage{bm}

\begin{document}

\title{Flexible Non-intrusive Dynamic Instrumentation for WebAssembly}

\author{Ben L. Titzer}
\email{btitzer@andrew.cmu.edu}
\affiliation{%
	\institution{Carnegie Mellon University}
	\city{Pittsburgh}
	\state{PA}
	\country{USA}}

\author{Elizabeth Gilbert}
\email{evgilber@andrew.cmu.edu}
\affiliation{%
	\institution{Carnegie Mellon University}
	\city{Pittsburgh}
	\state{PA}
	\country{USA}}

\author{Bradley Wei Jie Teo}
\email{bradleyt@andrew.cmu.edu}
\affiliation{%
	\institution{Carnegie Mellon University}
	\city{Pittsburgh}
	\state{PA}
	\country{USA}}

\author{Yash Anand}
\email{yashanan@andrew.cmu.edu}
\affiliation{%
	\institution{Carnegie Mellon University}
	\city{Pittsburgh}
	\state{PA}
	\country{USA}}

\author{Kazuyuki Takayama}
\email{ktakayam@andrew.cmu.edu}
\affiliation{%
	\institution{Carnegie Mellon University}
	\city{Pittsburgh}
	\state{PA}
	\country{USA}}

\author{Heather Miller}
\email{hmiller2@andrew.cmu.edu}
\affiliation{%
	\institution{Carnegie Mellon University}
	\city{Pittsburgh}
	\state{PA}
	\country{USA}}

\date{}

\thispagestyle{empty}





\newcommand{\oursys}{Wizard\xspace}
\newcommand{\ourjit}{Wizard's JIT\xspace}
\newcommand{\ourlang}{Virgil\xspace}
\newcommand{\mcode}{\texttt{M-code}\xspace}
\newcommand{\mstate}{\texttt{M-state}\xspace}
\newcommand{\mcodesub}[1]{\texttt{M-code}$_#1$\xspace}
\newcommand\API[1]{\texttt{#1}}
\newcommand\wasm[1]{\texttt{\textbf{#1}}\xspace}
\newcommand\wasms[1]{\texttt{\textbf{#1}s}}
\newcommand{\othersys}[1]{\texttt{\textbf{#1}}}
\newcommand\blfootnote[1]{%
	\begingroup
	\renewcommand\thefootnote{}\footnote{#1}%
	\addtocounter{footnote}{-1}%
	\endgroup
}
\newcommand{\boldplus}{$\bm{+}$}
\newcommand{\boldminus}{$\bm{-}$}

\begin{abstract}
    A key strength of managed runtimes over hardware is the ability to gain detailed insight into the dynamic execution of programs with instrumentation.
	Analyses such as code coverage, execution frequency, tracing, and debugging, are all made easier in a virtual setting.
	As a portable, low-level bytecode, WebAssembly offers inexpensive in-process sandboxing with high performance.
	Yet to date, Wasm engines have not offered much insight into executing programs, supporting at best bytecode-level stepping and basic source maps, but no instrumentation capabilities.
	In this paper, we show the first non-intrusive dynamic instrumentation system for WebAssembly in the open-source Wizard Research Engine\blfootnote{\texttt{https://github.com/titzer/wizard-engine}}.
	Our innovative design offers a flexible, complete hierarchy of instrumentation primitives that support building high-level, complex analyses in terms of low-level, programmable probes.
	In contrast to emulation or machine code instrumentation, injecting probes at the bytecode level increases expressiveness and vastly simplifies the implementation by reusing the engine's JIT compiler, interpreter, and deoptimization mechanism rather than building new ones.
	Wizard supports both dynamic instrumentation insertion and removal while providing consistency guarantees, which is key to composing multiple analyses without interference.
	We detail a fully-featured implementation in a high-performance multi-tier Wasm engine, show novel optimizations specifically designed to minimize instrumentation overhead, and evaluate performance characteristics under load from various analyses.
	This design is well-suited for production engine adoption as probes can be implemented to have no impact on production performance when not in use.
\end{abstract}

\maketitle
\pagestyle{plain}


\section{Introduction}
Programs have bugs and sometimes run slow.
Understanding the dynamic behavior of programs is key to debugging, profiling, and optimizing programs so that they execute correctly and efficiently.
Program behavior can be extremely complex with millions of interesting events~\cite{SideEffectRewriting}~\cite{DynamicMetricsForJava}~\cite{DynPrefetch}~\cite{RuntimeObjProfiler}~\cite{JPortal}.
Clearly manual inspection cannot scale and programmatic observation is required.

\subsection{Monitors and \mcode}

Just as we use the common term \emph{application} to refer to a self-contained program, we will use the term \emph{monitor} to refer to a self-contained \emph{analysis} which monitors an application and observes execution events and internal states.
For example, a profile monitor might count the number of runtime iterations of a loop or the execution count of basic blocks.
A monitor may \emph{instrument} programs using various mechanisms, before or during execution, execute runtime logic during program execution, and generate a post-execution report.
Of particular interest is a monitor's additional runtime storage and logic, which we refer to as \emph{monitor data} and \emph{monitor code}.
For example, a profile monitor's data includes counters and the monitor code includes the updates and reporting of counters.
Monitor code may be written in a high-level language and compiled to a lower form.
We will use \mcode to refer to the actual monitor code that will be executed at runtime.
\mcode may take many forms, including injected bytecode, source code, or machine code, utilities in the engine itself such as tracing modes, or extensions to the engine.

While most monitors aim to observe program behavior without changing it, the implementation technique may not always guarantee this.
For example, injecting \mcode by overwriting machine code in a native program is fraught with peril because native programs can observe machine-level details such as reading their own code as data.
Robustly separating monitor data from program data is a common problem.
Some approaches, such as emulation, avoid these low-level complexities and are inherently \emph{side-effect free} as they fully virtualize the execution model and \mcode can operate outside of this virtual execution context.

\subsubsection{Intrusive approaches}\label{sec:intrusive-approaches}

We say a monitor is \emph{intrusive} if it alters program behavior in a semantically observable way, i.e. it has side-effects on program state.
Intrusiveness is a property of the monitor together with the chosen technique for implementing instrumentation.
For example, instrumenting a native program that reads its own code as data could be intrusive if done with code injection, but non-intrusive with an emulator.
The intrusiveness of a monitor is independent of whether it \emph{perturbs} performance characteristics such as execution time and memory consumption\footnote{Often, non-intrusive implementation techniques allow measuring memory consumption of the program and monitor separately.}.

Instrumentation implementations that risk intrusiveness include static and dynamic code rewriting techniques.

\textbf{Static rewriting}.
If the execution platform does not directly offer debugging and inspection services, \emph{static rewriting} is often used where source code, bytecode, or machine code is injected directly into the program before execution.
Static rewriting has its advantages:
\vspace{2pt}
\begin{itemize}[leftmargin=10px]
	\item[\boldplus] no support from the execution platform is necessary; will work anywhere
	\item[\boldplus] not limited by the instrumentation capabilities of underlying execution platform; can do anything
	\item[\boldplus] inserted \mcode can be small and inline; approaches minimal overhead
	\item[\boldplus] instrumentation overhead is fixed before runtime; no dynamic instrumentation costs
\end{itemize}
\vspace{2pt}

However, static rewriting can also have disadvantages:
\begin{itemize}[leftmargin=10px]
	\item[\boldminus] \mcode intrudes on the state space of the code under test; easier to break the program
	\item[\boldminus] for machine- and bytecode-level instrumentation, \mcode must necessarily be low-level; more tedious to implement
	\item[\boldminus] for machine code, the binary may need to be reorganized to fit instrumentation; may not always be possible
	\item[\boldminus] offline instrumentation must instrument all possible events of interest; cannot dynamically adapt
	\item[\boldminus] source-level locations and mappings are altered by added code; additional mapping needed
	\item[\boldminus] \mcode perturbs performance in subtle and potentially complex ways; unpredictable performance impacts
	\item[\boldminus] pervasive instrumentation could massively increase code size; binary bloat
	\item[\boldminus] some information is only dynamically discoverable; could miss libraries, indirect calls, and generated code
\end{itemize}
\vspace{2pt}

Given these properties, this approach is frequently used and is demonstrated in the source-level tool Oron~\cite{Oron}, the bytecode-level tools BISM~\cite{BISM} and Wasabi~\cite{Wasabi}, and the machine-code-level tool EEL~\cite{EEL}, among others.

\textbf{Dynamic rewriting}.
In contrast to static rewriting, \emph{dynamic rewriting} allows a monitor to add \mcode at runtime.
This remedies some of the static rewriting disadvantages:
\vspace{2pt}
\begin{itemize}[leftmargin=2em]
	\item[\boldplus] can discover information only available at runtime
	\item[\boldplus] can instrument 100\% of the code
	\item[\boldplus] does not require recompilation or relinking of binary
	\item[\boldplus] potentially less code bloat
	\item[\boldplus] can dynamically adapt to program behavior, instrumenting more or less
	\item[\boldplus] implementation technique may be able to preserve some of the original addresses
\end{itemize}
\vspace{2pt}

However, it can have its own disadvantages:
\vspace{2pt}
\begin{itemize}[leftmargin=2em]
	\item[\boldminus] more instrumentation cost is paid at runtime
	\item[\boldminus] may make the execution platform vastly more complicated, e.g. requiring dynamic recompilation
	\item[\boldminus] the monitor is heavily coupled to the framework used to implement dynamic instrumentation
\end{itemize}
\vspace{2pt}

This approach is very common and is demonstrated in the source-level tool Jalangi~\cite{Jalangi}, the bytecode-level tool DiSL~\cite{DISL}, the machine-code-level tools Dyninst~\cite{Dyninst}, Pin~\cite{Pin} and Dtrace~\cite{DTrace}, among others (see Section~\ref{sec:related-work}).

\subsubsection{Non-intrusive approaches}\label{sec:nonintrusive-approaches}

Several techniques exist that do not alter the program code or behavior; they are side-effect free as their logic runs non-intrusively outside of the program space.
Debuggers for native binaries can use hardware-assisted techniques such as debug registers, JTAG, and process tracing APIs to debug programs directly on a CPU.
Emulators can support debugging and tracing easily in their interpreters.

Typically non-intrusive native mechanisms are slow, imposing orders of magnitude execution time overhead.
Yet accurately profiling high-frequency events that happen millions of times per second (branches, method calls, loops, and memory accesses) requires a more high-performance mechanism.
For limited cases such as profiling, Linux Perf~\cite{LinuxPerf} is a non-intrusive sampling profiler that can analyze programs directly running on the CPU.
Valgrind~\cite{Valgrind} and QEMU~\cite{QEMU} are emulators with analysis features and use dynamic binary translation via JIT compilation to reduce overheads.
Intel CPUs support a tracing mode known as Intel Processor Trace~\cite{IntelPT} that emits a densely-encoded history of branches, which can be used to reconstruct program execution paths.

Managed runtime environments offer high-performance implementations of languages and bytecode.
Many offer APIs for observing and interacting with a running program.
Support for instrumentation can be standard tracing modes, APIs for bytecode injection, or hot code reload (i.e. swapping the entire code of a function or class at a time).
For example, the Java Virtual Machine offers JVMTI~\cite{JvmTI}, and the .NET platform offers the .NET profiling API~\cite{DotNetProfiling}.
JVMTI offers both intrusive (dynamic bytecode rewriting with \texttt{java.lang.instrument}) and non-intrusive (\texttt{Agents}~\footnote{In JVMTI, users can write \emph{Agents} in native code that fire when \emph{Events} occur in the running application.
They then interface with the program to query state or control the execution itself.}) mechanisms.

The flexible sensor network simulator Avrora~\cite{Avrora}, powered by a microcontroller emulator, allows monitors to attach \mcode to code and memory locations as well as clock events~\cite{NonIntrusive}.
The \mcode is written in Java and therefore runs outside of the emulated CPU.
Its cycle-accurate interpreter runs microcontroller code faster than realtime on desktop CPUs without the need for a JIT.

Emulator and VM-based approaches still have the disadvantages of dynamic rewriting, but have more advantages:
\begin{itemize}[leftmargin=2em]
	\item[\boldplus] non-intrusive instrumentation; side-effect free
	\item[\boldplus] simplifies source-level address mapping
	\item[\boldplus] no need to reorganize binaries
	\item[\boldplus] can reuse existing JIT in engine; no instrumentation-specific JIT
	\item[\boldplus] does not require writing \mcode in a low-level language
	\item[\boldplus] does not perturb memory consumption; logical separation between program and monitor memory
\end{itemize}

\subsection{WebAssembly}

WebAssembly~\cite{WasmPldi}, or Wasm for short, is a portable, low-level bytecode that serves as a compilation target for many languages, including C/C++, Rust, AssemblyScript, Java, Kotlin, OCaml, and many others.
Initially released for the Web, it has since seen uptake in many new contexts such as Cloud~\cite{CloudflareWasm} and Edge computing~\cite{WasmEdgeDancer,FastlyEdge}, IoT~\cite{WasmIotOs,Aerogel}, and embedded and industrial~\cite{WasmIndMach} systems.
Wasm is gaining momentum as the primary sandboxing mechanism in many new computing platforms, as its execution model robustly separates Wasm module instance state.
The format is designed to be load- and run-time efficient with many high-performance implementations.
Wasm comes with strong safety guarantees, starting with a strict formal specification~\cite{WasmSpec}, a mechanically-proven sound type system~\cite{WasmMechSpec}, and implementations being subjected to verification~\cite{WasmSandboxing}.

Yet to date, no standard APIs for Wasm instrumentation exist.
Only intrusive instrumentation techniques exist today.
To work around the lack of standard APIs for instrumentation, several static bytecode rewriting tools have emerged~\cite{Wasabi}~\cite{AspectWasm}.

Wasm engines achieve near-native performance through AOT or JIT compilation.
Compilation is greatly simplified (over dynamic binary translation) as Wasm's code units are modules and functions rather than unstructured, arbitrarily-addressable machine code.
While Wasm JITs give excellent performance, some engines such as JavaScriptCore~\cite{JavaScriptCore} and wasm3~\cite{Wasm3} employ interpreters either for startup time or memory footprint.
Interpreters also help debuggability and introspection; recent work~\cite{FastWasmInterp} outlined a fast in-place interpreter design in the \oursys Research Engine.

\subsection{Our contributions}

\textbf{Flexible Non-Intrusive Instrumentation.}
In this work, we describe the first dynamic, non-intrusive (side-effect free) instrumentation framework for WebAssembly and detail its implementation in the open-source \oursys~\cite{WizardEngine} Research engine \footnote{\texttt{https://github.com/titzer/wizard-engine}}.
We show how to implement efficient support for \emph{probes} in a multi-tier Wasm engine and how to build useful, complex analyses from this basic building block, including tracing, profiling, and debugging.

\textbf{Consistent Dynamic Instrumentation.}
In contrast with prior work, our system also supports the dynamic insertion \emph{and removal} of individual probes.
For example, Pin supports dynamic clearing of instrumentation on a region of code, but it is not probe-specific.
Further, we make consistency guarantees about when insertion and removal take effect, which allows multiple analyses to be seamlessly composed.

\textbf{Zero Overhead When Not Used.}
This framework imposes zero overhead for disabled instrumentation.
To our knowledge, it is the first system that leverages \emph{dispatch table switching} to implement global probes, \emph{bytecode overwriting} for local probes, and specific JIT compiler support to achieve zero overhead in all execution tiers.

\textbf{JIT Intrinsification.}
We further show novel JIT optimizations that reduce the overhead of common instrumentation tasks by intrinsifying some probes.
We evaluate the effectiveness on a standard suite of benchmarks and place our system's performance in context with related work.

\textbf{Engine Mechanism Reuse.}
In contrast to Pin~\cite{Pin} and DynamoRIO~\cite{DynamoRIO}, our work makes only minor additions to the existing execution tiers of the Wasm engine (a few hundred lines of code), rather than a new, purpose-built JIT (tens of thousands of lines of code).
In the Wizard multi-tier research engine, we cleverly\footnote{We observe that the deoptimization (and on-stack replacement) mechanism already solves the hard problem of correct transfer between optimization levels and can be repurposed for instrumentation, saving tons of code.} reuse its deoptimization mechanism to achieve these consistency guarantees without needing to build a custom mechanism or resort to interpretation only.

\textbf{Feasible Production Adoption.}
Together, these innovations make it feasible for production engines to provide direct support for instrumentation without adding unnecessary complexity, putting powerful capabilities into the hands of application developers.

\begin{figure*}
	\includegraphics[width=7.1in]{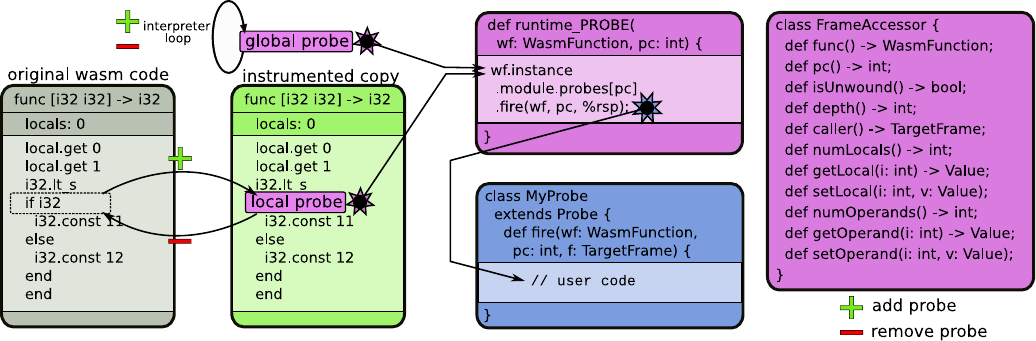}
	\caption{Illustration of instrumentation in the interpreter. Global probes can be inserted into the interpreter loop and local probes are implemented via bytecode overwriting. The \API{FrameAccessor} API allows a probe programmatic access to the state in the Wasm frame.}
	\label{fig:int-instrumentation}
\end{figure*}

\section{Non-intrusive instrumentation in \oursys}

High-performance virtual machines optimize execution time by cheating.
JIT compiler optimizations skip some unobservable execution steps of the abstract machine.
For example, not every update of a local variable or operand stack value is modeled at runtime, but function-local storage is virtualized and register-allocated.
Yet monitoring a program for dynamic analysis inherently observes intermediate states of a program, rather than just its final outcome.
Dynamic analyses typically observe states of the abstract machine, so VMs that support introspection must materialize the abstract states whenever requested.
For example, a dynamic analysis can observe any of the function-level storage such as the local variables and operand stack.

\textbf{Where to instrument programs?}
Most monitors instrument code to observe the flow of execution or program data.
With code instrumentation, locations in the original program code (e.g. bytecode offset, address, line number) become the natural points of reference.
This makes it intuitive to use an instrumentation API to attach \mcode to program locations which will fire when that point is reached during execution.

\textbf{Monitoring in \oursys with probes.}
A dynamic analysis for \oursys involves writing a \API{Monitor} in \ourlang~\cite{VirgilPldi} against an engine API.
With this API, everything about a Wasm program's execution can be observed \emph{on demand}, including any/every bytecode executed, any/every internal state computed, and all interaction with the environment.
Monitors observe execution by inserting \emph{probes} that fire callbacks before specified events or states occur.
Callbacks are dynamic logic but their \mcode can be statically compiled into the engine.
Their \mcode is efficient machine code that the engine invokes directly from either the interpreter or JIT-compiled code.
Since this \mcode executes as part of the engine and the engine virtualizes Wasm program state, monitors are inherently non-intrusive.

\textbf{Probes are maximally general.}
While many systems~\cite{TuningFork,JPortal,ShadowVM,Roadrunner} offer event traces that can be analyzed asynchronously or offline, probes are more fundamental since they enable the insertion of arbitrary code at arbitrary locations.
For example, probes can \emph{generate} event traces or react to program behavior, but event traces do not offer the ability to influence execution, an inherent capability of synchronous probes.
Thus, we say that probes are \emph{complete} in the sense that every type of instrumentation can be built from them.

Figure~\ref{fig:int-instrumentation} illustrates the probe hooks offered by \oursys and their implementation in the interpreter.

\subsection{Global Probes}

The simplest type of probe is a \emph{global probe}, which fires a callback \emph{for every instruction executed by the program}.
Clearly global probes are \emph{complete} since they can execute arbitrary logic at any point in execution.
Despite their inefficiency, global probes are still useful.
A global probe is the easiest way to implement tracing, counting, or the \texttt{step-instruction} operation of a debugger.

Global probes are easy to implement in an interpreter; its main loop or dispatch sequence simply contains a check for any global probe(s) and calls them at each iteration.
They also are the slowest \mcode because, even with a JIT, they effectively reduce the VM to an interpreter\footnote{E.g. in a compile-only engine, a compilation mode which inserts a call to fire global probes before \emph{every} instruction suffices, but bloats generated code and has marginal performance benefit over an interpreter.}.

Unfortunately, the simple implementation technique of an extra check per interpreted instruction imposes overhead even when not enabled, which tempts VMs to have different production and debug builds.
A key innovation in \oursys (Section~\ref{sec:optimization}) is to implement global probes with \emph{dispatch table switching}, which imposes zero overhead when global probes are not enabled, obviating the need for a separate debug build.
This technique also allows efficient dynamic insertion and removal of global probes, which we have found to be a useful mechanism for implementing some analyses, shown in Section~\ref{sec:after-instruction}.
Regardless of implementation efficiency, an engine can achieve instrumentation-completeness by adding only global probe support.

\subsection{Local Probes}

Many dynamic analyses are \emph{sparse}, only needing to instrument a subset of code locations.
For this reason, \oursys also allows \emph{local probes} to be attached to specific locations in the bytecode.
At runtime, the engine fires local probes just \emph{before} executing the respective instruction.
Each Wasm instruction can be identified uniquely by its module, function, and byte offset from the start of the function, making the triple \API{(module, funcdecl, pc)} a natural location identifier in the API.
Since local probes only fire when reaching a specific location, they are more convenient for implementing analyses such as branch profiling, call graph analysis, code coverage, breakpoints, etc.

Local probes can be significantly more efficient than global probes for several reasons:
\begin{itemize}[leftmargin=2em]
	\item zero overhead for uninstrumented instructions
	\item efficient implementation in interpreter and compilers
	\item compilers can optimize around local probes
\end{itemize}
\vspace{4pt}

Like global probes, local probes are \emph{complete}; both can be implemented in terms of each other at the cost of efficiency\footnote{Emulating local probes with global probes can be done with logic that looks up each local probe in \mstate, and global probes can be emulated by inserting local probes everywhere, but incurs overhead from data structure lookups in the engine.}.

\subsection{The \API{FrameAccessor} API}

\newcommand{\accessor}{\API{FrameAccessor}\xspace}

While many analyses need only the sequence of program locations executed, more advanced dynamic analyses like taint tracking, fuzzing and debugging observe program states.
To allow probe callbacks access to program state, they receive not only the program location, but also a lazily-allocated object with an API for reading state, called the \emph{\accessor}.

The \accessor provides callbacks a fa\c{c}ade~\cite{DesignPatterns} with methods to read frame state, abstracting over the machine-level details of frames, which often differ between execution tiers and engine versions.
They offer a stable interface to a frame where, due to dynamic optimization and deoptimization, the engine may change the frame representation during the execution of a function.

A \accessor object represents a single stack frame and is allocated when a callback first requests state other than the easily-available \API{WasmFunction} and \API{pc}.
Importantly, the identity of this object is observable to probes so that they can implement higher-level analyses across multiple callbacks.
At the implementation level, execution frames maintain the mapping to their accessor by storing a reference in the frame itself, called the \emph{accessor slot}.
The slot is not used in normal execution, but imposes a one machine word space overhead; its execution time impact should be negligible.

\textbf{Stackwalking and callstack depth.}
The \accessor API allows walking up the callstack to callers so monitors can implement context-sensitive analyses and stacktraces.
The depth of the call stack alone is also often useful for tracing or context-sensitive profiling, so \accessor objects include a \API{depth()} method, which a VM can implement slightly more efficiently.

\textbf{Dangling accessor objects.}
\accessor objects are allocated in the engine's state space (for \oursys, the managed heap), and since probes are free to store references to them across multiple callbacks, it is possible that the accessor object outlives the execution frame that it represents\footnote{With ownership, as in Rust, lifetime annotations can statically prevent a \accessor object from escaping from a \emph{single} callback, yet some monitors legitimately want to track frames across multiple callbacks.}.
While the accessor object itself will be eventually reclaimed, it is problematic if \mcode accesses frames that have been unwound.
We identified a number of implementation mechanisms to protect the runtime system from buggy monitors.

Possible solutions include:
\begin{enumerate}[leftmargin=2em]
	\item \label{clear} \textbf{Clear accessor on entry}.
	      Upon entry to a Wasm function, the accessor slot in the execution frame is unconditionally set to \texttt{null}.
	\item \textbf{Invalidate accessor on return}.
	      A dynamic check is performed on all returns from a function; if the accessor slot points to a valid \accessor object, the object itself is invalidated (e.g. by setting a field in the object to \texttt{false}).
	\item \textbf{Invalidate accessors on unwind}.
	      When unwinding frames for a trap or exception thrown, which is typically done in the runtime rather than compiled code, the accessor object itself is invalidated.
	\item \label{guards} \textbf{Return guards invalidate accessor}.
	      When an accessor slot is set, the return address for the frame is also redirected to a trampoline that will invalidate the accessor object before returning to the actual caller.
	\item \label{check} \textbf{\accessor methods check frame validity}.
	      Every call to an accessor object checks that the underlying machine frame points back at the accessor object.
	\item \textbf{\accessor methods check self validity}.
	      Every call to an accessor object checks the object's validity field.
\end{enumerate}

Our solution is to minimize checks in the interpreter and compiled code and favor checks at the \accessor API boundary and corresponds to a combination of \ref{clear}, \ref{guards}, and \ref{check}.
This relies on stack frame layout invariants: function entry clears the accessor slot, the first request for the \accessor materializes the object, and subsequent accessor calls compare the accessor slot to a cached stack pointer in the object.
To make these checks bulletproof to monitor bugs, {\accessor}s should be invalidated on return, e.g. with a runtime check\footnote{Or a return guard trampoline, which avoids any runtime overhead.}.

\subsection{Consistency guarantees}\label{subsec:consistency-guarantees}

Many analyses can be implemented by making use of dynamic probe insertion and removal.
Other analyses, particularly debuggers, could make modifications to frames that alter program behavior.
When do new probes and frame modifications take effect?
Providing consistency guarantees is a key innovation in our system that makes composing multiple analyses reliable.
With these guarantees, probes from multiple monitors do not interfere, making monitors \emph{composable} and \emph{deterministic}.
Monitors can be used in any combination without explicit foresight in their implementation.

\subsubsection{Deterministic firing order}

What should happen if a probe $p$ at location $L$ fires and inserts another probe $q$ at the same location $L$?
Should the new probe $q$ also fire before returning to the program, or not?
Similarly, if probes $p$ and $q$ are inserted on the same event, is their firing order predictable?

We found that a guaranteed probe firing order is subtly important to the correctness of some monitors (e.g. the function entry/exit utility shown in Section~\ref{sec:func_entry}).
For this reason, we guarantee three \emph{dynamic probe consistency} properties:
\newcommand\property[1]{\underline{\textbf{#1}}}
\begin{itemize}[leftmargin=2em]
	\item \property{Insertion order is firing order}: Probes inserted on the same event $E$ fire in the same order as they were inserted.
	\item \property{Deferred inserts on same event}: When a probe fires on event $E$ and inserts new probes on the same $E$, the new probes do not fire until the next occurrence of $E$.
	\item \property{Deferred removal on same event}: When a probe fires on event $E$ and removes probes on the same $E$, the removed probes \emph{do} fire on this occurrence of $E$ but not subsequent occurrences.
\end{itemize}

\subsubsection{Frame modifications}

As shown, the \accessor provides a mostly \emph{read-only} interface to program state.
Since monitors run in the engine's state space, and not the Wasm program's state space, by construction this guarantees that monitors do not alter the program behavior.
However, some monitors, such as a debugger's \texttt{fix-and-continue} operation, or fault-injection, intentionally change program state.

For an interpreter, modifications to program state, such as local variables, require no special support, since interpreters typically do not make assumptions across bytecode boundaries.
For JIT-compiled code, any assumption about program state could potentially be violated by \mcode frame modifications.
Depending on the specific circumstance, continuing to run JITed code after state changes might exhibit unpredictable program behavior\footnote{
	True even for baseline compilers like \oursys's compiler, which perform limited optimizations like register allocation and constant propagation.}.

It's important for the engine to provide a consistency model for state changes made through the \accessor.
When monitors explicitly \emph{intend} to alter the program's behavior, it is natural for them to expect state changes to take effect immediately, \emph{as if} the program is running in an interpreter.
Thus, our system guarantees:

\vspace{4pt}
\begin{itemize}[leftmargin=2em]
	\item \property{Frame modification consistency}: State changes made by a probe are immediately applied, and execution after a probe resumes with those changes.
\end{itemize}
\vspace{4pt}

This effectively requires immediate deoptimization of a frame, also guaranteed by JVMTI.
Otherwise, if execution continues in JIT-compiled code, almost any invariant the JIT relied on could be invalid, and it may appear that updates have not occurred yet, violating consistency.

\subsubsection{Multi-threading}

While \oursys is not currently multi-threaded, WebAssembly does have proposals to add threading capabilities which \oursys must eventually support.
That brings with it the possibility of multi-threaded instrumentation.
Locks around insertion and removal of probes should maintain our consistency guarantees through serializing dynamic instrumentation requests.
Our design inherently separates monitor state from program state.
Thus data races on the monitor state are the responsibility of the monitors, for example by using lock-free data structures and/or locks at the appropriate granularity.
The \API{FrameAccessor} can also include synchronization to prevent data races on Wasm state\footnote{Note: frames are by-definition thread local; races can only exist if the monitor itself is multi-threaded and \accessor objects are shared racily.}.

\subsection{Function Entry/Exit Probes}\label{sec:func_entry}

Probes are a low-level, instruction-based instrumentation mechanism, which is natural and precise when interfacing with a VM.
Yet many analyses focus on function-level behavior and are interested in calls and returns.
Instrumentation hooks for function entry/exit make such analyses much easier to write.

At first glance, detecting function entry can be done by probing the first bytecode of a function, and exit can be detected by probing all \wasms{return}, \wasms{throw}, and \wasms{br} that target the function's outermost block.
However, some special cases make this tricky.
First, a function may begin with a \wasm{loop}; the entry probe must distinguish between the first entry to a function, a backedge of the loop, and possible (tail-)recursive calls.
Second, local exits are not enough: frames can be unwound by a callee throwing an exception caught higher in the callstack.

Should the VM support function entry/exit as special hooks for probes?
Interestingly, we find this is not strictly necessary.
This functionality can be built from the programmability of local probes and offered as a library.
There are several possible implementation strategies: 1) use entry probes that push the \accessor objects onto an internal stack, with exit probes popping; 2) sampling the stack depth via the \accessor's \API{depth()} method; or 3) by instrumenting, and thus ignoring, loop backedges.
Thus, function entry/exit reside above global/local probes in the hierarchy of instrumentation mechanisms.
This is further evidence that the programmability of probes allows building higher-level instrumentation utilities for more expressive dynamic analyses.

\subsection{After-instruction}
\label{sec:after-instruction}

Some analyses, such as branch profiling or dynamic call graph construction, are naturally expressed as \mcode that should run \emph{after} an instruction rather than \emph{before}.
For example, profiling which functions are targets of a \wasm{call\_indirect} would be easiest if a probe could fire \emph{after} the instruction is executed and a frame for the target function has been pushed onto the execution stack.
However, the API has no such functionality.

Should the VM support an ``after-instruction'' hook directly?
Interestingly, we find that like function entry/exit, the unlimited programmability of probes allows us to invoke \mcode \emph{seemingly after} instructions.
For example, suppose we want to execute probe $p$ after a \wasm{br\_table} (i.e. Wasm's switch instruction).
We identified at least three strategies:

\begin{itemize}[leftmargin=2em]
	\item A probe $q_p$ executed \emph{before} the \wasm{br\_table} can use the \accessor object to read the top (\wasm{i32} value) of the operand stack, determine where the branch will go, and dynamically insert probe $p$ at that location.

	\item Insert probes into all targets of the \wasm{br\_table}.
          Since \wasm{br\_table} has a fixed set of targets, we can insert probes once and use \mstate to distinguish reaching each target from the \wasm{br\_table} versus another path.
          This only works in limited circumstances; other instructions like \wasm{call\_indirect} have an unlimited set of targets.
	\item Insert a global probe for just one instruction and remove it after.
The probe will fire on the next instruction, wherever that is, then the probe will remove itself.
For a use case like this, it's important that dynamically enabling global probes doesn't ruin performance, e.g. by deoptimizing all JIT-compiled code.
We show in Section~\ref{sec:dispatch-table} how dispatch-table switching can make this use case efficient.
\end{itemize}

With multiple strategies to emulate its behavior, an after-instruction hook resides above global/local probes in the instrumentation mechanism hierarchy.

\section{The Monitor Zoo}\label{sec:monitor-types}

The wide variety and ease with which analyses are implemented\footnote{Most monitors required a dozen or two lines of instrumentation code; in fact, most lines are usually spent on making pretty visualizations of the data!} showcases the flexibility of having a fully programmable instrumentation mechanism in a high-level language.
Users activate monitors with flags when invoking \oursys (e.g. \texttt{wizeng -\xspace-monitors=MyMonitor}), which instrument modules at various stages of processing before execution and may generate post-execution reports.
Examples of monitors we have built include a variety of useful tools.

The \textbf{Trace monitor} prints each instruction as it is executed.
While many VMs have tracing flags and built-in modes that may be spread throughout the code, \oursys already offers the perfect mechanism: the global probe.
Instruction-level tracing in \oursys simply uses one global probe.
Other than a short flag to enable it, there is nothing special about this probe; it uses the standard \accessor API as it prints instructions and the operand stack.

The \textbf{Coverage monitor} measures code coverage.
It inserts a local probe at every instruction (or basic block), which, when fired, sets a bit in an internal datastructure and then \emph{removes itself}.
By removing itself, the probe will no longer impose overhead, either in the interpreter or JITed code.
Eventually, all executed paths in the program will be probe-free and JITed code quality will asymptotically approach zero overhead.
This is a good example of a monitor using dynamic probe removal.

The \textbf{Loop monitor} counts loop iterations.
It inserts \API{CountProbe}s at every loop header and then prints a nice report.
This is a good example of a counter-heavy analysis.

The \textbf{Hotness monitor} counts every instruction in the program.
It inserts \API{CountProbe}s at every instruction and then prints a summary of hot execution paths.
Another example of a counter-heavy analysis.

The \textbf{Branch monitor} profiles the direction of all branches.
It instruments all \wasm{if}, \wasm{br\_if} and \wasm{br\_table} instructions and uses the top-of-stack to predict the direction of each branch.
It is a good example of non-trivial \accessor usage.

The \textbf{Memory monitor} traces all memory accesses.
It instruments all loads and stores and prints loaded and stored addresses and values.
Another good example of non-trivial \accessor usage.

The \textbf{Debugger REPL} implements a simple read-eval-print loop that allows interactive debugging at the Wasm bytecode level.
It supports breakpoints, watchpoints, \texttt{single-step}, \texttt{step-over}, and changing the state of value stack slots.
It primarily uses local probes but uses a global probe to implement single-step functionality.
This monitor is a good example of dynamic probe insertion and removal.
It is also the only monitor (so far) that modifies frames.

The \textbf{Calls monitor} instruments callsites in the program and records statistics on direct calls and the targets of indirect calls.
Its output can be used to build a dynamic call graph from an execution.

The \textbf{Call tree profiler} measures execution time of function calls and prints self and nested time using the full calling-context tree.
It can also produce flame graphs.
It inserts local probes at all direct and indirect callsites and all return locations~\footnote{\oursys has preliminary support for the proposed Wasm exception handling mechanism, but does not yet have monitoring hooks for unwind events.}.
It is a good example of a monitor that measures non-virtualized metrics like wall-clock time.

\section{Optimizing probe overhead}\label{sec:optimization}
\begin{figure*}
	\includegraphics[width=7.1in]{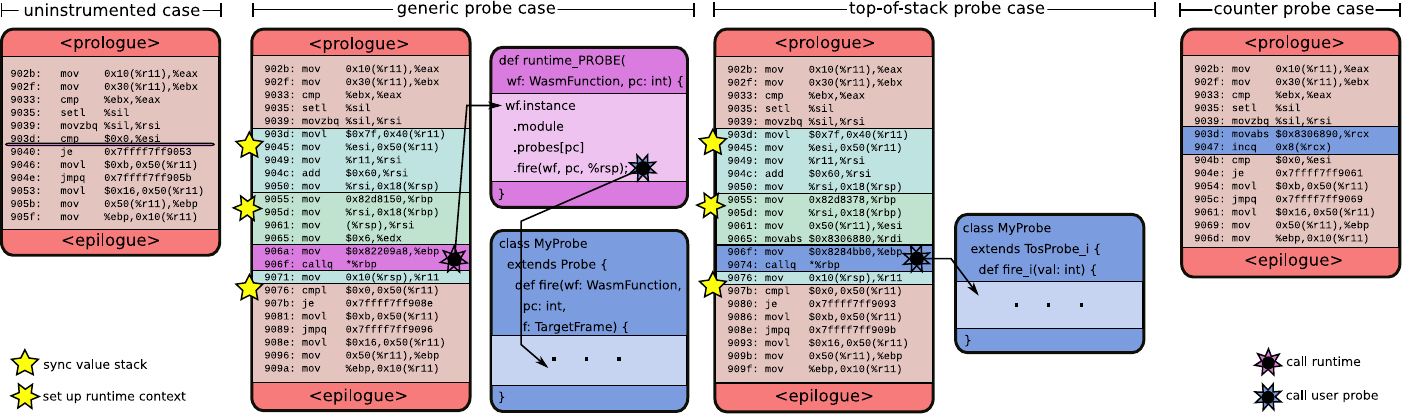}
	\caption{Code generated by \oursys's baseline JIT for different types of \mcode implemented with probes.
		The machine code sequence for generic probes is more general than for probes that only need the top-of-stack value, versus a fully-intrinsified counter probe.}
	\label{fig:jit-instrumentation}
\end{figure*}

Optimizations in \oursys's interpreter and JIT compiler reduce overhead for both global and local probes, see the effectiveness of this technique in Section~\ref{subsec:jit-optimization-evaluation}.
We define \emph{overhead} as the execution time spent in neither application code nor \mcode, but in transitions \emph{between} application and \mcode or additional work in the runtime system and compiler.

\subsection{Optimizing global probes in the interpreter}
\label{sec:dispatch-table}
Global probes, being the most heavyweight instrumentation mechanism, are supported only in the interpreter.
It is straightforward to add a check to the interpreter loop that checks for any global probes at each instruction.
However, this naive approach imposes overhead on all instructions executed, even if global probes are not enabled.
One option to avoid overhead when global probes are disabled is to have two different interpreter loops, one with the check and one without, and dynamically switch between them.
This comes at some VM code space cost, since it duplicates the entire interpreter loop and handlers.
Another approach described in~\cite{FastWasmInterp} avoids the code space cost by maintaining a pointer to the \emph{dispatch table} in a hardware register.
When global probes are not in use, this register points to a ``normal'' dispatch table without instrumentation; inserting a global probe switches the register to point to an ``instrumented'' dispatch table where all (256) entries point to a small stub that calls the probe(s) and then dispatches to the original handler via the ``normal'' dispatch table.
Both code duplication and dispatch-table switching are suitable for production, as they allow the VM to support global probes while imposing no overhead when disabled.

Dynamically adding and removing global probes shouldn't ruin performance, as they might be used to implement ``after-instruction'' or to trace a subset of the code, such as an individual function or loop.
Our design further extends~\cite{FastWasmInterp} by supporting global probes without deoptimizing JITed code.
This can be done by temporarily returning to the interpreter in the global probe mode.
In global probe mode, a different dispatch table is used, which, in addition to calling probes for every instruction, can use special handlers for certain bytecodes.
For example, the \wasm{loop} bytecode does not check for dynamic tier-up (which would cause a transfer to JITed code), \wasm{call} instructions reenter the interpreter (rather than entering the callee's JITed code, if any), and \wasm{return} returns only to the interpreter (rather than the caller's JIT code).
Otherwise, JIT code remains in-place.
Removing global probes leaves this mode and JIT code will naturally be reentered as normal.
See Section~\ref{subsec:strategies-for-multi-tier-consistency} for how we guarantee consistency after state modifications.
To our knowledge, our design is the first to support switching into a heavyweight instrumentation mode and back without discarding any JITed code, preserving performance.

\subsection{Optimizing local probes in the interpreter}

Both \oursys's interpreter and baseline JIT support local probes.
In the interpreter, local probes impose no overhead on non-probed instructions by using in-place bytecode modifications.
With \emph{bytecode overwriting}, inserting a local probe at a location $L$ overwrites its original opcode with an otherwise-illegal \wasm{probe} opcode.
The original unmodified opcode is saved on the side.
When the interpreter reaches a \wasm{probe} opcode, the Wasm program state (e.g. value stack) is already up-to-date; it saves the interpreter state, looks up the local probe(s) at the current bytecode location, and simply calls that \mcode callback.
This is somewhat reminiscent of machine code overwriting, a technique sometimes used to implement debugging or machine code instrumentation (Pin, \texttt{gdb} and DynamoRIO).
However, our approach is vastly simpler and more efficient as it doesn't require hardware traps or solving a nasty code layout issue \textemdash only a single bytecode is overwritten.

In \oursys, since the callback is compiled machine code, the overhead is a small number of machine instructions to exit the interpreter context and enter the callback context.
After returning from \mcode, $L$'s original opcode is loaded (e.g. by consulting an unmodified copy of the function's code) and executed.
Removing a probe is as simple as copying the original bytecode back; the interpreter will no longer trip over it.
In contrast, Pin allows disabling by removing \emph{all instrumentation} from a specified region of the original code, which effectively reinstalls the original code, an all-or-nothing approach rather than having control at the probe granularity.
Overwriting has two primary advantages over bytecode injection; the original bytecode offsets are maintained, making it trivial to report locations to \mcode, and insertion/removal of probes is a cheap, constant-time operation.
Consistency is trivial; the bytecode is always up-to-date with the state of inserted instrumentation.

\subsection{Local probes in the JIT}

In a JIT compiler, local probes can be supported by injecting calls to \mcode into the compiled code at the appropriate places.
Since probe logic could potentially access (and even modify) the state of the program through the \accessor, a call to unknown \mcode must checkpoint the program and VM-level state.
For baseline code from \ourjit, the overhead is a few machine instructions more than a normal call between Wasm functions\footnote{Primarily because the calling convention models an explicit value stack.}.
Compilation speed is paramount to a baseline compiler, and bytecode parsing speed actually matters.
Similar to the benefits to interpreter dispatch, bytecode overwriting avoids any compilation speed overhead because the \wasm{probe} opcode marks instrumented instructions and additional checks aren't needed.
Overall, supporting probes adds little complexity to the JIT compiler; in \ourjit, it requires less than 100 lines of code.

\subsection{JIT intrinsification of probes}

While probes are a fully-programmable instrumentation mechanism to implement unlimited analyses, there are a number of common building blocks such as counters, switches, and samplers that many different analyses use.
For logic as simple as incrementing a counter every time a location is reached, it is highly inefficient to save the entire program state and call through a generic runtime function to execute a single increment to a variable in memory.
Thus, we implemented optimizations in \ourjit to \emph{intrinsify} counters as well as probes that access limited frame state.

Figure~\ref{fig:jit-instrumentation} shows how \oursys's baseline JIT optimizes different kinds of probes.
At the left, we have uninstrumented code.
For the generic probe case, the JIT inserts a call to a generic runtime routine calls the user's probe.
For the next more specialized case, the top-of-stack, it inserts a direct call to the probe's \API{fire} method, passing the top-of-stack value, skipping the runtime call overhead and the cost of reifying an expensive \accessor object.
In general values from the frame can be directly passed from the JITed code to \mcode.
Lastly, for the counter probe, we see that \ourjit simply inlines an increment instruction to a \emph{specific} \API{CountProbe} object without looking it up.

Other systems allow building custom inline \mcode.
For example, Pin offers using a type of macro-assembler that builds IR that it compiles into the instrumented program, which is very low-level, tedious, and error-prone.

\subsection{Monitor consistency for JITed code}

We just saw how a JIT can inline \mcode into the compiled code.
However, \mcode can change as probes can be inserted and removed during execution, making compiled code that has been specialized to \mcode out-of-date.
This problem can be addressed by standard deoptimization techniques such as on-stack-replacement back to the interpreter and invalidating relevant machine code.
To our knowledge, no prior bytecode-based system has employed deoptimization to support dynamic instrumentation of an executing frame, but offer only hot code replacement.

\subsection{Strategies for multi-tier consistency}\label{subsec:strategies-for-multi-tier-consistency}

There are several different strategies for guaranteeing monitor consistency in a multi-tier engine like \oursys.
We identified four plausible strategies:

\begin{enumerate}[leftmargin=2em]
	\item When instrumentation is enabled, disable the JIT.
	\item When instrumentation is enabled, disable only \emph{relevant} JIT optimizations.
	\item Upon frame modification, recompile the function under different assumptions about frame state and perform on-stack-replacement from JITed to JITed code.
	\item Upon frame modification, perform on-stack-replacement from JITed code to the interpreter.
\end{enumerate}

Strategy 1) is the simplest to implement for engines with interpreters, but slow.
A production Wasm engine could achieve functional correctness and the key consistency guarantees at little engineering cost, leaving instrumented performance as a later product improvement.
Strategy 2) eliminates interpreter dispatch cost, but, ironically, is actually a lot of work in practice, since it introduces modes into the JIT compiler and optimizations must be audited for correctness.
The compiler becomes littered with checks to disable optimization and ultimately the JIT emits very pessimistic code.
Strategy 3) has other implications for JIT compilation, such as requiring support for arbitrary OSR locations\footnote{Most JITs that allow tier-up OSR into compiled code only do so at loop headers.}, which is also significant engineering work.

In \oursys, we chose strategy 4, which we believe to be not only the simplest, but most robust.
Frame modifications trigger immediate deoptimization of \emph{only the modified frame}\footnote{We observe that the JIT-compiled code for a function is not \emph{invalid}, it is only the state of the single frame that now differs from assumptions in the JIT code. New calls to the involved function can still legally enter the existing JIT code.}, rewriting it in place to return to the interpreter.
In the \emph{dynamic} tiering configuration mode, sending an execution frame back to the interpreter due to modification doesn't banish it there forever; if it remains hot, it can be recompiled under new assumptions\footnote{Pathological cases can occur where hot frames are repeatedly modified, constantly transferring between interpreter and JITed code. A typical fix employed in many VMs is to simply limit the number of times a function can be optimized and offer user diagnostics.}.
This means frame modification support requires the interpreter; \oursys will not allow modifications in the JIT-only configuration.

Inserting or removing probes in a function also triggers deoptimization of JITed code for the function and sends existing frames back to the interpreter.
This is different than a frame modification, because the JIT may have specialized the code to instrumentation at the time of compilation; the code is actually invalid w.r.t. the instrumentation it should execute.
Like with frame modifications, hot functions will eventually be recompiled.
It's likely that such highly dynamic instrumentation scenarios would perform better by using \mstate to enable and disable their probes rather than repeatedly inserting and removing them, which confounds engine tiering heuristics.

\begin{figure}
	\includegraphics[width=0.5\textwidth]{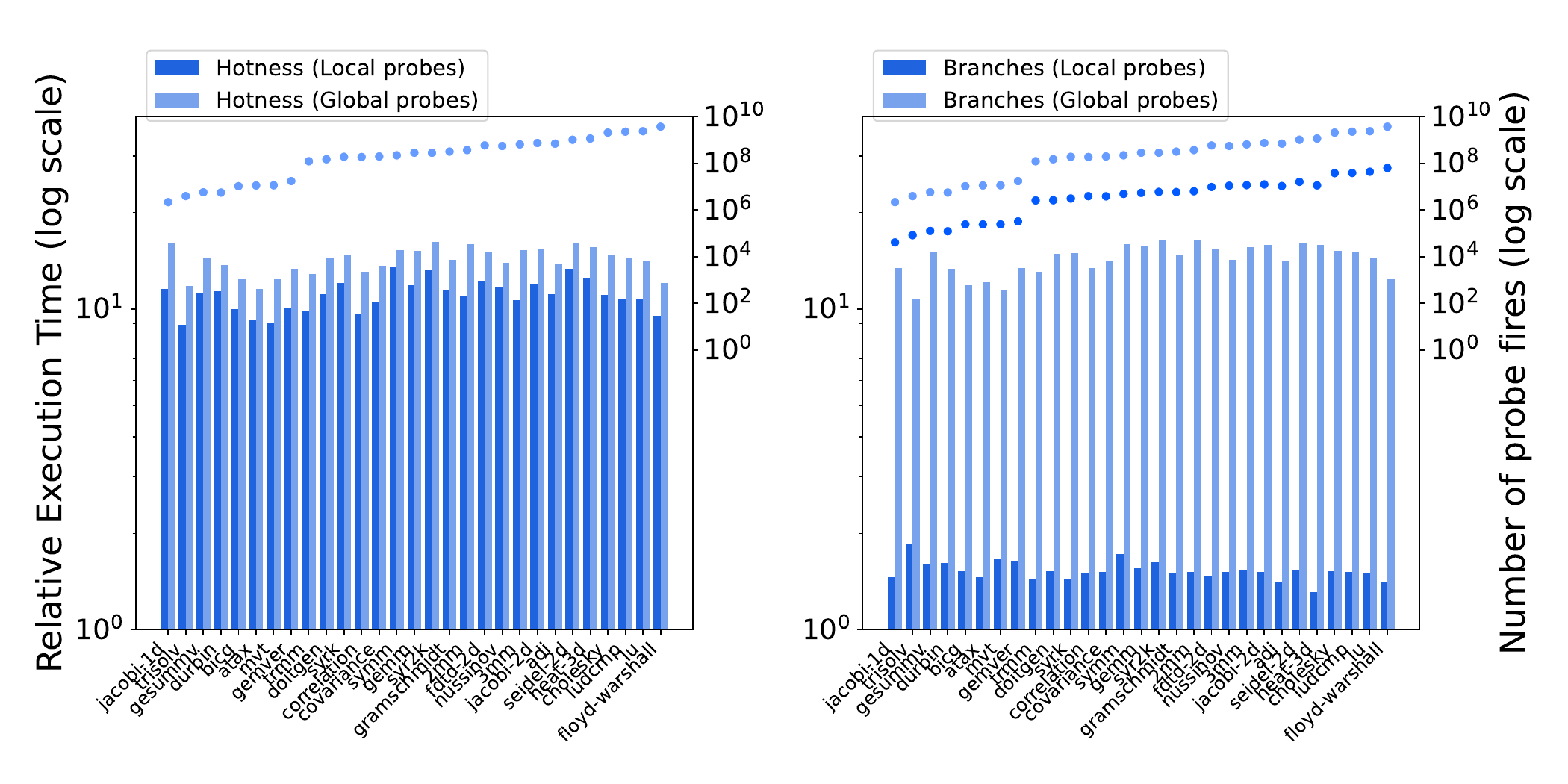}
	\caption{Average relative execution time for the hotness monitor (left) and branch monitor (right), when implemented with local probes and when implemented with a global probe on the PolyBenchC suite. Points above the bars denote number of probe fires.}
	\label{fig:experiment-local-global}
\end{figure}

\section{Evaluation}

In this section, we evaluate performance of monitoring code using three suites of benchmarks and several different implementation strategies.
We compare instrumenting Wasm code in \oursys, bytecode rewriting, bytecode injection with Wasabi, and native code instrumentation with DynamoRIO.

\subsection{Evaluation setup}
We evaluate the performance of \oursys by executing Wasm code under both the interpreter and JIT using different monitors and measure total execution time of the entire program, including engine startup and program load.
We chose the ``hotness'' and ``branch'' monitors (described in Section \ref{sec:monitor-types}).
The hotness monitor instruments every instruction\footnote{Obviously, it is more efficient to count basic blocks. We chose to count every instruction in order to \emph{maximize} instrumentation workload.} with a local \API{CountProbe}, which is representative of monitors with many simple probes.
The branch monitor probes branch instructions and tallies each destination by accessing the top of the operand stack.
Compared to the hotness monitor, probes in the branch monitor are more sparse but more complex.

These monitors were chosen because they strike a balance between being powerful enough to capture insights about the execution of a program, yet simple enough to be implemented in other systems.
They are also likely to instrument a nontrivial portion of program bytecode.

\textbf{Benchmark Suites.}
We run Wasm programs from three benchmark suites: PolyBench/C~\cite{PolyBench} with the \texttt{medium} dataset, Ostrich~\cite{Ostrich} and Libsodium~\cite{Libsodium} and average execution time over 5 runs.

Given instrumented execution time $T_i$ and uninstrumented execution time $T_u$, we define \emph{absolute overhead} as the quantity $T_i-T_u$ and \emph{relative execution time} as the ratio $T_i/T_u$.
We report relative execution time for \oursys's interpreter, \ourjit (with and without intrinsification), DynamoRIO, Wasabi, and bytecode rewriting in Figures \ref{fig:experiment-diff-frameworks-all} and \ref{fig:experiment-diff-frameworks-aggr}.

\subsection{Global vs local probes}

Global probes can emulate the behavior of local probes, but impose a greater performance cost by introducing checks at every bytecode instruction.
We compare two implementations of the branch and hotness monitors, one using a global probe and the other using local probes.
Both are executed in \oursys's interpreter, since \ourjit doesn't support global probes.
The results can be found in Figure \ref{fig:experiment-local-global}.
For the hotness monitor, since the number of probe fires is the same for local and global probes, the relative overhead is similar across all programs.
For the branch monitor, local probes on branch instructions have relative execution times between $1.0$--$2.2\times$, whereas it is between $7.7$--$16.4\times$ for global probes. 

\begin{figure}
	\includegraphics[width=0.5\textwidth]{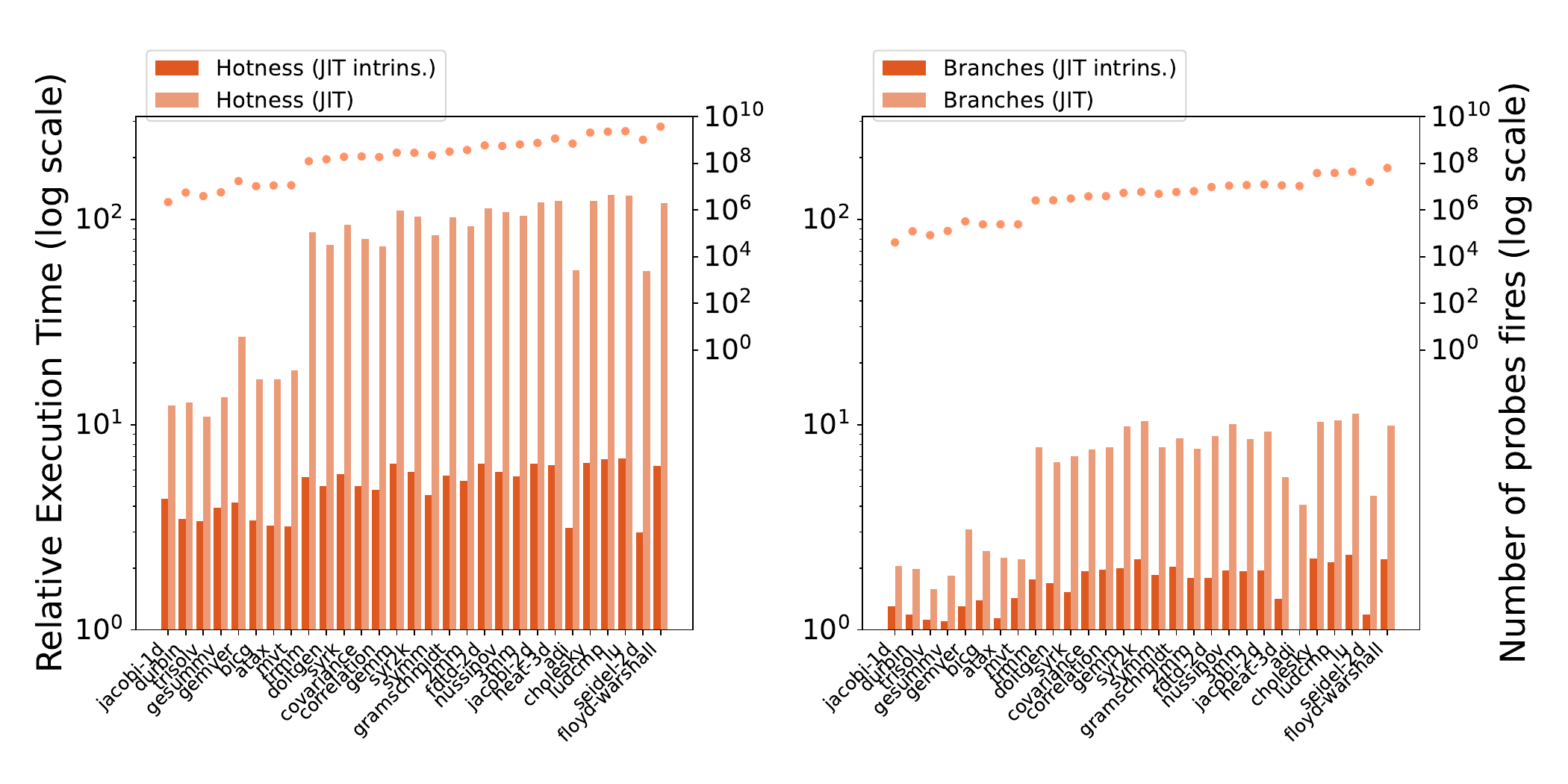}
	\caption{Average relative execution times for the hotness (left) and branch monitors (right), with and without probe intrinsification on the PolyBenchC suite. Ratios are relative to uninstrumented JIT execution time. Points above the bars denote number of probe fires.}
	\label{fig:experiment-jit-intrinsic}
\end{figure}

\subsection{JIT optimization of count and operand probes}\label{subsec:jit-optimization-evaluation}
Section \ref{sec:optimization} describes how \ourjit intrinsifies some types of probes to reduce overhead.
We evaluate JIT intrinsification in Figure \ref{fig:experiment-jit-intrinsic} and report the relative execution time of instrumented over uninstrumented execution.

For the hotness monitor, which counts the execution frequency of every instruction,
we observed relative execution times between $7$--$134\times$. 
This is due to the high cost of switching between JIT code and the engine at every instruction.
With intrinsification, the same monitor has relative execution times between $2.2$--$7.7\times$. 

We performed a similar experiment to evaluate the effectiveness of JIT intrinsification of top-of-stack operand probes by measuring execution times with the branch monitor.
We see that intrinsification improves relative execution times from $1.0$--$16.6\times$ with the base JIT to $1.0$--$2.8\times$. 
The improvement is smaller for branch probes than for \API{CountProbes}, because a call into the probe's \mcode remains, whereas \API{CountProbes} are fully inlined (see Figure \ref{fig:jit-instrumentation}).

\begin{figure}
	\begin{minipage}[t]{0.25\textwidth}
		\centering
		\includegraphics[width=\textwidth]{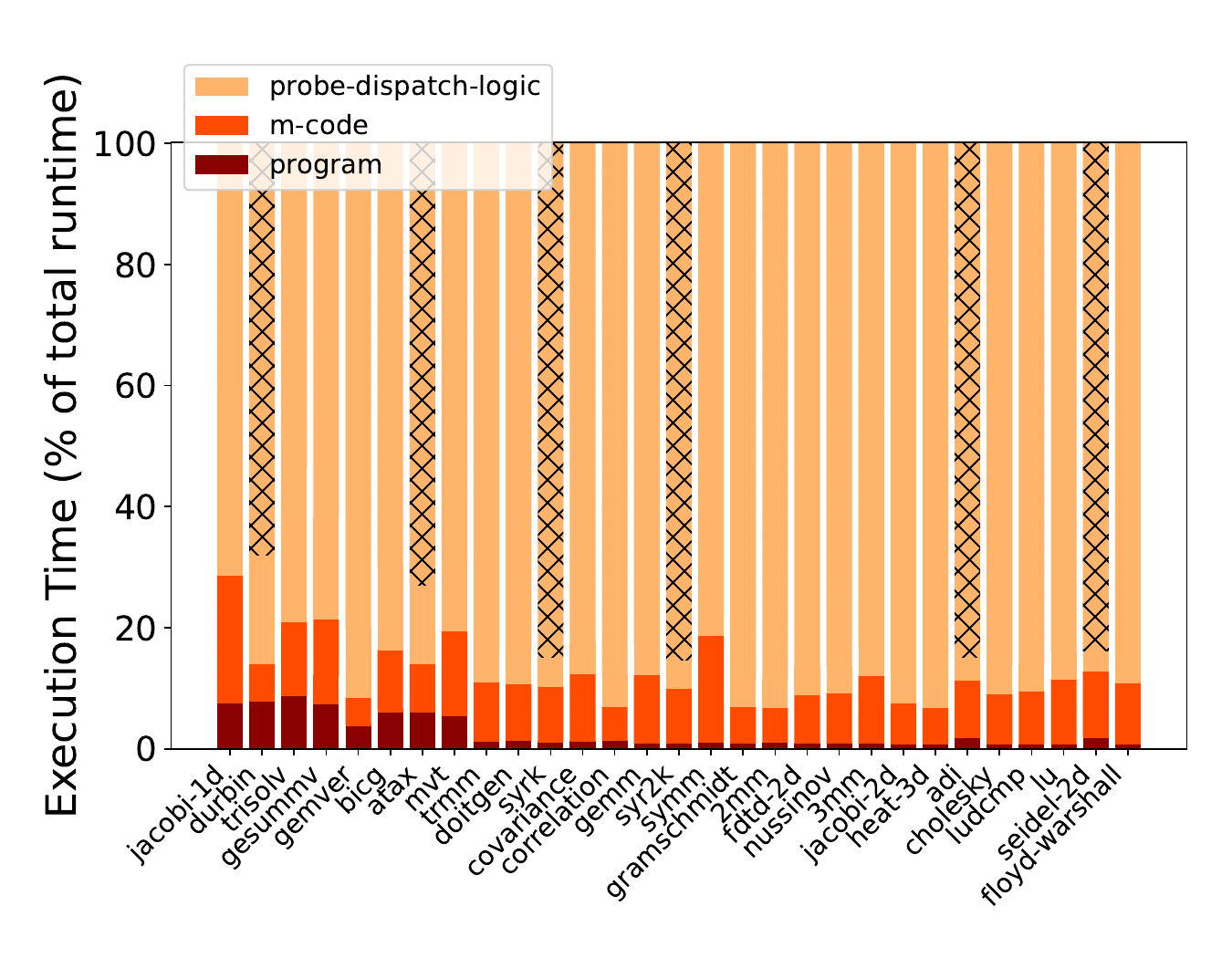}
	\end{minipage}%
	\begin{minipage}[t]{0.25\textwidth}
		\centering
		\includegraphics[width=\textwidth]{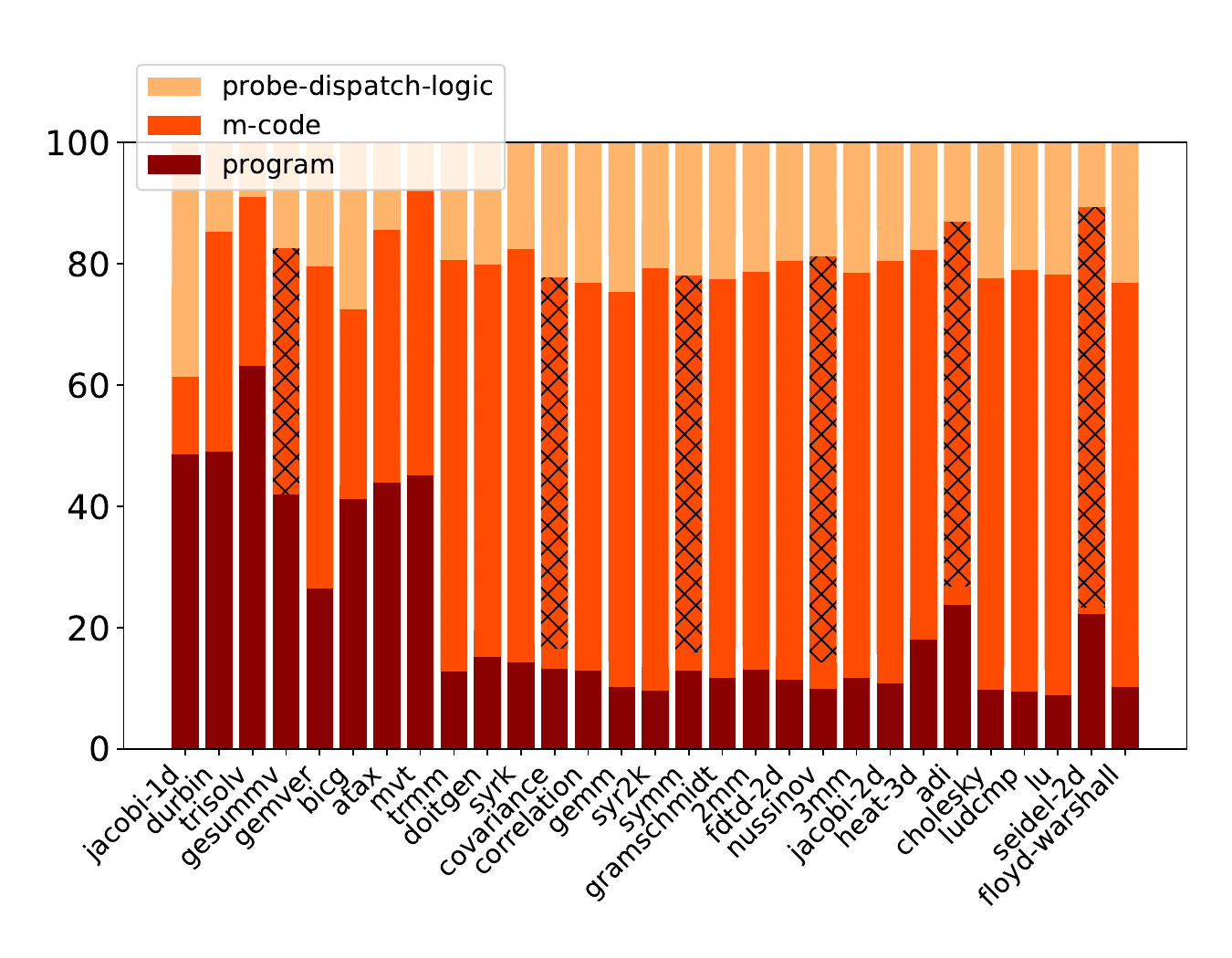}
	\end{minipage}
	\caption{Execution time decomposition of hotness (left) and branch monitors (right) into \mcode and probe dispatch overhead with and without probe intrinsification on the PolyBenchC suite. The cross-hatched regions represent overhead saved by intrinsification.}
	\label{fig:experiment-jit-intrinsic-decomp}
\end{figure}

We further decompose the runtime of the benchmarks into the time spent in the program's JIT-compiled code ($T_\text{JIT}$), time in \mcode ($T_M$), and time in the probe dispatch logic ($T_\text{PD}$).
This decomposition is done by recording: \begin{enumerate}[leftmargin=2em]
	\item The uninstrumented execution time of code in the JIT, which approximates $T_\text{JIT}$;
	\item The instrumented execution time with empty probes (probes with empty \API{fire} functions), which approximates $T_\text{PD}+T_\text{JIT}$;
	\item The instrumented execution time with actual probes, which gives $T_\text{PD}+T_M+T_\text{JIT}$.
\end{enumerate}

The results of this analysis for the branch and hotness monitors are in Figure \ref{fig:experiment-jit-intrinsic-decomp}.
Execution time without JIT intrinsification is shown as the entire bar for each program.
The cross-hatched portions of each bar represent the execution time saved by intrinsification.
For the non-intrinsified branch monitor, the overhead $T_\text{PD}+T_M$ is dominated by \mcode.
In the intrinsified case, the overhead is dominated by probe dispatch, and the \mcode overhead is reduced substantially: calling the top-of-stack operand probe's \mcode still requires significant spilling on the stack and a call, contributing to runtime overhead.
The \mcode overhead no longer includes time for construction of the \API{FrameAccessor} as it is not necessary.

As for the non-intrinsified hotness monitor, the overhead is dominated by the probe dispatch overhead as probes are simpler but fired more frequently.
In the intrinsified case, there is almost no \mcode overhead as counter probes do not have custom \API{fire} functions; the counter increment is entirely inlined.
The remaining probe dispatch overhead comes from the monitor setup and reporting.

\subsection{Interpreter vs. JIT}

\begin{figure*}
	\includegraphics[width=\textwidth]{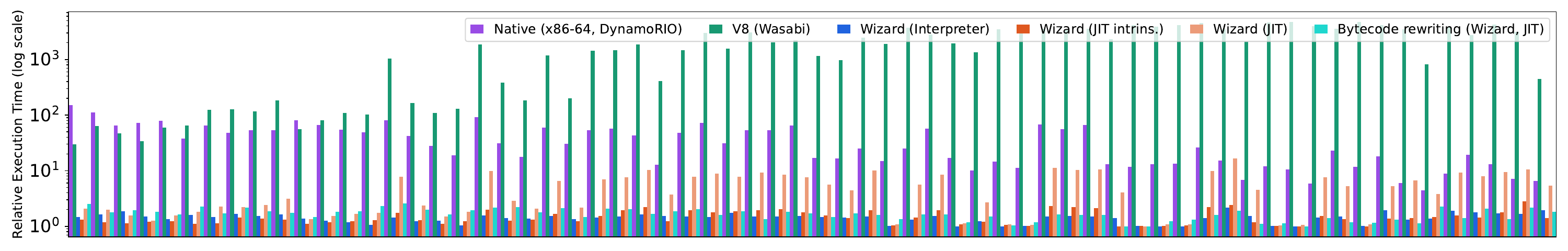}
	\includegraphics[width=\textwidth]{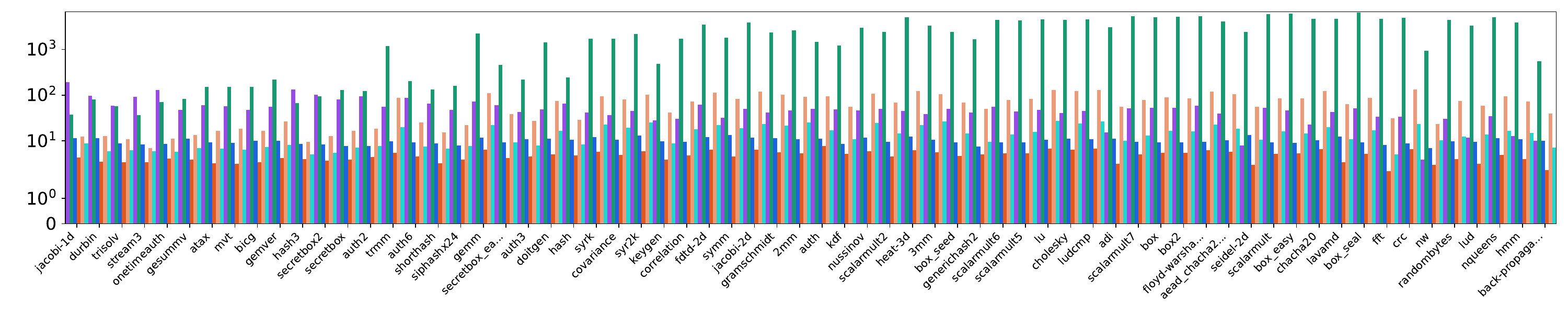}
	\caption{Relative execution times of the hotness monitor (bottom) and branch monitor (top) in \oursys, Wasabi, and DynamoRIO across all programs on all suites, sorted by absolute execution time. Ratios are relative to uninstrumented execution time.}
	\label{fig:experiment-diff-frameworks-all}
\end{figure*}

\begin{figure}
	\begin{minipage}[t]{0.25\textwidth}
		\centering
		\includegraphics[width=\textwidth]{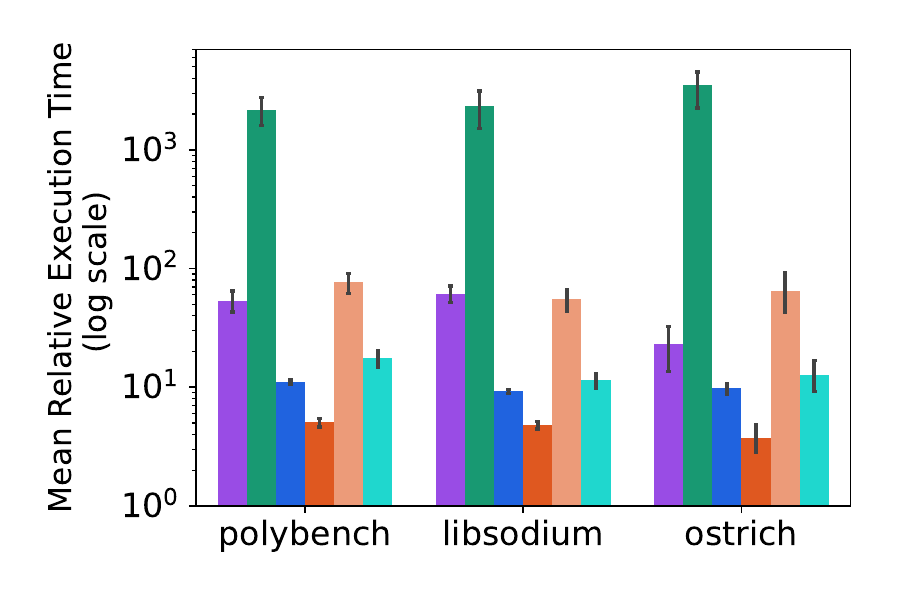}
	\end{minipage}%
	\begin{minipage}[t]{0.25\textwidth}
		\centering
		\includegraphics[width=\textwidth]{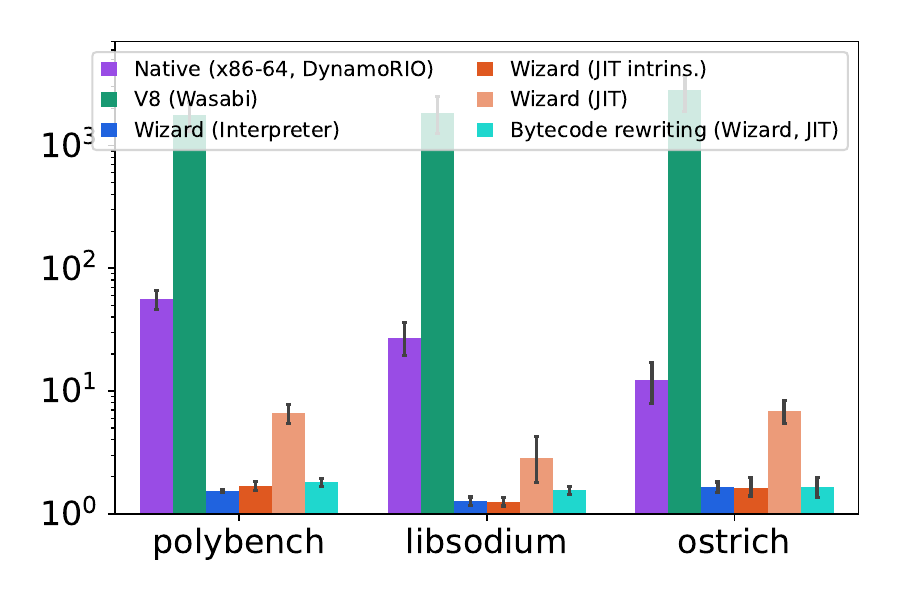}
	\end{minipage}
	\begin{minipage}[t]{0.25\textwidth}
		\centering
		\includegraphics[width=\textwidth]{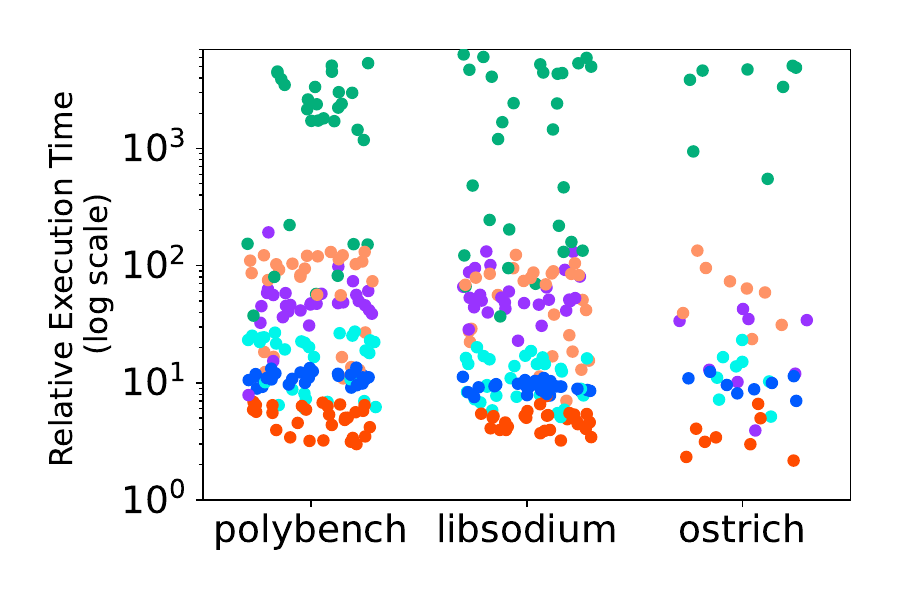}
	\end{minipage}%
	\begin{minipage}[t]{0.25\textwidth}
		\centering
		\includegraphics[width=\textwidth]{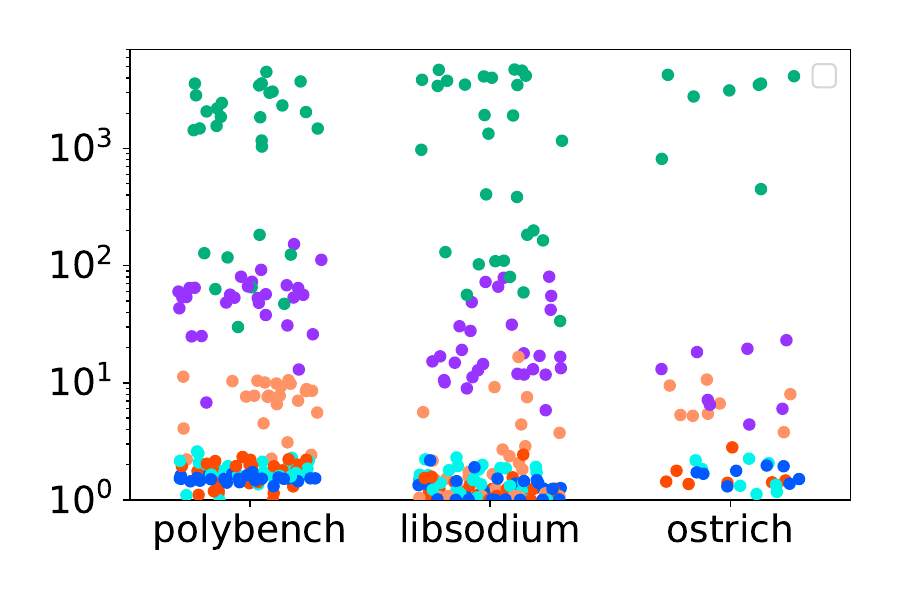}
	\end{minipage}
	\caption{Mean relative execution times of the hotness monitor (left) and branch monitor (right) in \oursys, Wasabi, and DynamoRIO across the three suites. Ratios are relative to uninstrumented execution time.}
	\label{fig:experiment-diff-frameworks-aggr}
\end{figure}

We find that the relative overhead of monitors running in \oursys's interpreter is much lower than the JIT, for two reasons: the interpreter runs much slower, and less additional work is done in checkpointing state.
In contrast, calls to local probes in the JIT require checkpointing to support the \accessor API.
Data in Figures \ref{fig:experiment-diff-frameworks-all} and \ref{fig:experiment-diff-frameworks-aggr} show that, for the branch monitor, the relative execution time in the interpreter is $1.0$--$2.2\times$ as compared to $1.0$--$16.6\times$ in \ourjit. 
In the higher-workload hotness monitor, this difference is exacerbated: the relative execution time in the interpreter is $7.0$--$13.5\times$ as compared to $7.0$--$134\times$ in the JIT. 
Although relative execution times differ substantially, absolute overhead between the two modes is comparable: for the branch monitor, the mean overhead in the interpreter is $2.6$s and $2.3$s in the JIT.

\subsection{Comparison with bytecode rewriting}

Bytecode rewriting is an example of static instrumentation described in Section \ref{sec:intrusive-approaches}.
Using Walrus \cite{Walrus}, a Wasm transformation library written in Rust, we implemented the hotness and branch monitors by rewriting bytecode \cite{wasm-bytecode-instrumenter}.
For the hotness monitor, we inject counting instructions before each instruction, and for the branch monitor, before each branching instruction.
Counters are stored in memory, necessitating loads and stores.
We evaluated the performance of the transformed Wasm bytecode when run in \ourjit, and compared it to their respective monitors in \oursys.
From Figure \ref{fig:experiment-diff-frameworks-aggr}, we observe that the intrinsified JIT execution time is lower than that of bytecode rewriting for both monitors.

\subsection{Comparison with Wasabi}

Wasabi is a dynamic instrumentation tool that runs analyses on Wasm bytecode using a JavaScript engine.
Since Wasabi instrumentation must be written in JavaScript, it requires a Wasm engine that also runs JavaScript, such as V8~\cite{V8}.
For this comparison, we use V8 in its default mode (two compiler tiers)\footnote{We also conducted experiments limiting V8 to its baseline compiler, for a closer comparison to \ourjit. Results indicate that the Wasm compiler makes little difference; the overhead is dominated by JavaScript execution and transitions between JavaScript. For more JIT comparisons see \cite{WizardJit}.}.
Figure \ref{fig:experiment-diff-frameworks-all} includes data for Wasabi on \othersys{v8}.
Wasabi instrumentation is vastly slower than \oursys instrumentation due to the overhead of calling JavaScript functions.
On average, a hotness monitor in Wasabi increases execution time $36.8$--$6350.2\times$, compared to $7$--$134\times$ for \ourjit (or $2.2$--$7.7\times$ with intrinsification).
The branch monitor also has a drastic performance impact of $29.9$--$4721.5\times$ in Wasabi, compared to $1.0$--$16.6\times$ for \ourjit (or $1.0$--$2.8\times$ with intrinsification). 

\subsection{Comparison with DynamoRIO}
We also compare with machine code instrumentation.
We cannot make a direct comparison, so instead, we compile the same benchmark programs to x86-64 assembly and instrument them with DynamoRIO, with analogous machine-code hotness and branch monitors.

The results are shown in Figures \ref{fig:experiment-diff-frameworks-all} and \ref{fig:experiment-diff-frameworks-aggr}.
Executing the native programs instrumented with a DynamoRIO hotness monitor is about $3.9$--$192\times$ slower than without instrumentation.
The hotness monitor has a substantial relative overhead because, among other things, DynamoRIO inserts instructions to spill and restore \texttt{EFLAGS} for each counter increment.
On the other hand, the DynamoRIO branch monitor slows execution time from $4.4$--$153\times$, again compared to $1.0$--$16.6\times$ for \ourjit (or $1.0$--$2.8\times$ with intrinsification).
This is likely because our DynamoRIO monitor is implemented with a function call at every basic block, which DynamoRIO can sometimes inline, since it works at the machine code level.
Its default inlining heuristics seem to give rise to unpredictable overheads.

\subsection{Evaluation summary}
We evaluated \oursys's instrumentation overheads by measuring the relative execution times of a branch and a hotness monitor across multiple standardized benchmarks and instrumentation approaches.
For monitors with sparse probes, like the branch monitor, local probes improve on the performance of global probes (relative execution time of $1.0$--$2.2\times$ versus $7.7$--$16.4\times$).
Running monitors in \ourjit further improves this performance to a relative execution time of $1.0$--$16.6\times$ without intrinsification and $1.0$--$2.8\times$ with intrinsification.
This greatly outperforms DynamoRIO and Wasabi, with relative execution times of $4.4$--$153\times$ and $29.9$--$4721.5\times$ respectively.
Surprisingly, JIT intrinsification can produce instrumentation overhead even lower than intrusive bytecode rewriting, shown in Figure \ref{fig:experiment-diff-frameworks-aggr}.
Our results show \oursys's instrumentation architecture is both flexible and efficient.

\vspace{-7pt}
\section{Related work}\label{sec:related-work}

Techniques for studying program behavior have been the subject of a vast amount of research.
They differ in \emph{when} to instrument (statically or dynamically), at \emph{which level} (source, IR, bytecode, or machine code), and \emph{what} mechanism is used to do so.
A key issue is that the analysis and the mechanism work together to analyze a program's behavior \emph{intrusively} or \emph{non-intrusively}, yet some mechanisms make it easier to implement non-intrusive analyses.

\vspace{-7pt}
\subsection{Code injection}
A common approach is to inject \mcode directly into programs~\cite{BinRewriteSurvey}, either statically or dynamically.
Injecting code into a program can be done \emph{inline} (directly inserted into code, often requiring binary reorganization), with \emph{trampolines} (jumps to out-of-line instrumentation code), or both.

\textbf{Static.}
Early tools for Java static bytecode instrumentation include Soot~\cite{Soot} and Bloat~\cite{NystromThesis}.
Later, with the rise of Aspect-Oriented Programming (AOP), tools emerged to target \emph{joinpoints}, such as DiSL~\cite{DISL}, AspectJ~\cite{AspectJ} and BISM~\cite{BISM}.
Oron~\cite{Oron} reduced the performance overhead of JavaScript source-level instrumentation by targetting AssemblyScript and compiling the instrumented program to Wasm for execution.
For Wasm, tools are now emerging such as the aspect-oriented ~\cite{AspectWasm}, and Wasabi~\cite{Wasabi}, which injects trampolines into Wasm bytecode that call instrumentation code provided as JavaScript.

\textbf{Dynamic.}
FERRARI~\cite{FERRARI} statically instruments core JDK classes while dynamically instrumenting all others using \texttt{java.lang.instrument}.
SaBRe~\cite{SaBRe} injects instrumentation at load-time, thus paying the rewriting cost once at startup rather than continuously during execution.
DTrace~\cite{DTrace,DTrace2}, inspired by Paradyn~\cite{Paradyn} and other tools, enables tracing at both the user and kernel layer of the OS by operating inside the kernel itself and uses dynamically-injected trampolines.
Dyninst~\cite{Dyninst} interfaces with a program's CFG and maps modifications to concrete binary rewrites.
A user can tie \mcode to instructions or CFG abstractions (e.g. function entry/exit).
It can do this statically or at any point during execution and changes are immediate.
Recent research in this direction \cite{ToggleProbe} \cite{Odin} \cite{TowardMinimalMonitoring} focuses on reducing instrumentation overhead with a variety of low-level optimizations.

\subsection{Recompilation.}

Compiled programs can be \emph{recompiled} to inject code using several techniques.

\textbf{Static.}
Early examples of static lifting for instrumentation include ATOM~\cite{AtomTools}, followed by EEL~\cite{EEL} with finer-grained instrumentation.
Etch~\cite{Etch}, through observing an initial program execution, discovered dynamic program properties to inform static instrumentation.
Other examples include Vulcan~\cite{Vulcan}, which injects code into lifted Win32 binaries.

\textbf{Dynamic.}
Both DynamoRIO~\cite{DynamoRIO} and Pin~\cite{Pin} use dynamic recompilation of native binaries to implement instrumentation.
They differ somewhat on subtle implementation details, how \mcode is injected, and performance characteristics, but fundamentally work by recompiling machine code for a given ISA to the same ISA.
Their JIT compilers are purpose-built for instrumentation and are basic-block and trace-cache based.
They run code in the original process and reorganize binaries, and can be intrustive, particularly if \mcode is supplied as low-level native code.
We are not aware of strong consistency guarantees (Section~\ref{subsec:consistency-guarantees}) in the face of dynamically adding and removing instrumentation.

RoadRunner~\cite{Roadrunner} is a dynamic analysis system for Java based on event streams, primarily focused on race detection.
It uses a custom classloader to inject calls to instrumentation.
Analyses are formulated in terms of pipes and filters over event streams, allowing composability.
It offers some specific inlining optimizations that avoid the overhead of events in some circumstances.
Since the analysis code runs in the same state space and on the same threads, it can both perturb performance and alter concurrency characteristics of highly-multithreaded programs.

ShadowVM~\cite{ShadowVM} builds on JVMTI to provide non-intrusive instrumentation with low perturbation by running the monitor on a separate JVM and asynchronously processing events as they occur.
It is primarily suited for program observation, as it does not directly support state modifications.
On load, an instrumentation process dynamically inserts hooks through bytecode rewriting that trap to native code to asynchronously communicate event notifications to the monitor.
According to published material, ShadowVM does not support dynamically inserting and removing hooks during program execution.

\vspace{-7pt}
\subsection{Emulation}
QEMU~\cite{QEMU} is a widely-used CPU emulator that  virtualizes a user-space process while supporting non-intrusive instrumentation.
Valgrind~\cite{Valgrind}, primarily used as a memory debugger, is similar.
As emulators, both can run a guest ISA on a different host ISA, and both use JIT compilers to make emulation fast.
Thus, their JIT compilers are not necessarily ``purpose-built'' for instrumentation, but for cross-compilation.
Avrora~\cite{Avrora}, a microcontroller emulator and sensor network simulator, provides an API to attach \mcode to clock events, instructions, and memory locations.

\vspace{-7pt}
\subsection{Direct engine support}

Runtime systems can be designed with specific support for instrumentation.
In .NET~\cite{DotNetProfiling}, users build profiler DLLs that are loaded by the CLR into the same process as a target application.
The CLR then notifies the profiler of events occurring in the application through a callback interface.
The JVM Tool Interface~\cite{JvmTI} allows Java bytecode instrumentation and also \emph{agents} to be written against a lower-level internal engine API that supports attaching callbacks to events.
Examples of events are method entry and exit, but nothing as fine-grained as reaching individual bytecodes.
To assess the performance overhead of handling MethodEntry events, we wrote a Calls monitor using JVMTI in C.
When run on the famously indirect-call-heavy Richards benchmark it imposes $50$--$100\times$ overhead.
In contrast, for the same program compiled to Wasm and running with \oursys's Calls monitor, the overhead was measured to be $2.5$--$3\times$.

\vspace{-7pt}
\section{Conclusion and Future Work}

In this paper, we showed the first non-intrusive dynamic instrumentation framework for Wasm in a multi-tier Wasm research engine that imposes zero overhead when not in use.
Modifications to the interpreter and compiler tiers of \oursys are minimal; just a few hundred lines of code.
Novel optimizations reduce instrumentation overhead and perform well for sparse analysis and acceptably well for heavy analysis.
Our robust consistency guarantees make our system the first to support composing multiple analyses seamlessly.

While probes offer a complete instrumentation mechanism for code, many analyses instrument other events, such as memory accesses, traps, etc.
As we saw with function entry/exit and after-instruction hooks, libraries can implement higher-level hooks \emph{using} probes; but if directly supported by the engine, these hooks can be implemented more efficiently, e.g. hardware watchpoints for memory accesses.

In this work, we showed monitors written against \oursys's engine APIs in a high-level language.
Generic probes use runtime calls to compiled \mcode.
Massive speedups are possible from intrinsifying \emph{certain} probes by inlining all or part of their \mcode.
What if \mcode was instead supplied in an \emph{IR} the JIT could \emph{inline}?
We plan to explore \emph{Wasm bytecode} as just that IR.

\begin{acks}
  This work is supported in part by NSF Grant Award \#2148301, as well as funding and support from the WebAssembly Research Center.
  Thanks to Anthony Rowe and Arjun Ramesh for important discussions and comments on drafts of this work.
  Thanks to Sa\'ul Cabrera, Erin Ren, and Jeff Charles at Shopify, Ulan Degenbaev and Yan Chen at DFinity, and Chris Woods at Siemens.
\end{acks}

\bibliographystyle{plain}
\bibliography{paper}

\appendix
%
%
%
%

\section{Artifact Appendix}

\subsection{Abstract}

This artifact description contains information for how to reproduce all results in this paper.
We describe system requirements, how to set up an environment and run our scripts that produce data and exact figures in the paper, as well as how to modify the artifact to run your own custom experiments.
Our package contains all scripts, benchmarks, monitors, and engines used.
We also provided all of our results in the package so others can do direct data comparison.

\subsection{Artifact Meta-Information}
{\small
    \begin{itemize}[leftmargin=2em]
        \item \textbf{Benchmarks:} The following benchmarking suites are used in our experiments:
             \begin{itemize}
                 \item PolyBench/C~\cite{PolyBench} with the \texttt{medium} dataset, version 4.2.
                 \item Ostrich~\cite{Ostrich}, version 1.0.0.
                 \item Libsodium~\cite{Libsodium}, there are three different variations of the Libsodium benchmark as follows:
                 \begin{itemize}
                     \item \texttt{libsodium}, the base libsodium suite, version 0.7.13.
                     \item \texttt{libsodium-2021}, a variation pulled from the 2021-Q1 directory at \url{https://github.com/jedisct1/webassembly-benchmarks}.
                     \item \texttt{libsodium-no-bulk-mem}, a variation of base libsodium above without bulk memory operations.
                 \end{itemize}
              \end{itemize}
              All of these benchmark suites have been included in the \texttt{suites} directory of the artifact.
        \item \textbf{Compilation:} Since we provide the benchmarks compiled to Wasm, a Wasm compiler is unnecessary. However, a Rust compiler for the \texttt{wasm-bytecode-instrumenter} and Wasabi tools is necessary. See the next section for the required version of \texttt{rustc}.
        \item \textbf{Transformations:} For the bytecode rewriting experiment, we use Walrus (version 0.20.1), a Rust library for Wasm transformations, in our \texttt{wasm-bytecode-rewriter} to inject our Wasm instrumentation.
              This crate's repo is publicly available at: \url{https://github.com/rustwasm/walrus}.
              Wasabi also does Wasm transformations to inject calls, follow instructions in the \texttt{README.md} file at the root directory of the artifact for how to set up this tool.
        \item \textbf{Binaries:}
              We used Wizard~\cite{WizardEngine} version \texttt{23a.1617} for our experimentation.
              Since this version did not have support for enabling/disabling certain features via command line flags, we directly manipulated flags in the source code and compiled each required Wizard configuration for our experiments and made them available in the \texttt{bin} folder.
              The following describes each binary's configuration:
                \begin{itemize}
                    \item \texttt{base/wizeng.x86-64-linux}.
                          This is the base compilation of Wizard with no flags modified.
                    \item \texttt{local-global/wizeng.x86-64-linux}.
                          In order to use a monitor in Wizard, its code must be present in the binary.
                          We extended the base Wizard to contain both the local and global implementations of the hotness and branch monitors in this binary.
                    \item \texttt{fast-count/wizeng.x86-64-linux}.
                          This binary was compiled with the \texttt{intrinsifyCountProbe} and \break
                          \texttt{intrinsifyOperandProbe} flags enabled in the \break
                          \texttt{src/engine/Tuning.v3} file.
                    \item \texttt{empty-probes/wizeng.x86-64-linux}.
                          To reiterate, in order to use a monitor in Wizard, its code must be present in the binary.
                          We extended the base Wizard to contain variations of the hotness and branch monitors with no \mcode in their inserted probes.
                    \item \texttt{empty-probes-fast-count/wizeng.x86-64-linux}.
                          This binary is a combination of the above \texttt{fast-count} and \texttt{empty-probes} configurations.
                \end{itemize}
              We have also included the binary \texttt{btime} in the \texttt{bin} folder that calculates various types of timing characteristics of a program's execution.
        \item \textbf{Run-time environment:} This project must be run on an \texttt{x86\_64} Linux machine.
            It is not necessary to have \texttt{sudo} access as software can be installed/symlinked in a user's home directory.
            However, having \texttt{sudo} access would substantially simplify the installation process.
        \item \textbf{Metrics:} We report \emph{relative execution time} and \emph{absolute overhead} for \oursys's interpreter, \ourjit (with and without intrinsification), DynamoRIO, Wasabi, and bytecode rewriting.
              Given instrumented execution time $T_i$ and uninstrumented execution time $T_u$, we define \emph{absolute overhead} as the quantity $T_i-T_u$ and \emph{relative execution time} as the ratio $T_i/T_u$.
        \item \textbf{Output:} There are two different outputs for our scripts. The \texttt{experiment*.bash} scripts output CSV files. The \texttt{plot*.py} scripts output graphs as PDF and SVG files.
              We have provided our own results inside the \texttt{csv} and \texttt{figures} folders for comparison.
        \item \textbf{Experiments:} Follow the instructions in the \texttt{README.md} file at the base directory of the artifact for how to run experiments.
        \item \textbf{How much disk space required (approximately)?:} 10 GB
        \item \textbf{How much time is needed to prepare workflow (approximately)?:} We expect experiment preparation to take around 30 minutes (if all builds/installs work well).
        \item \textbf{How much time is needed to complete experiments (approximately)?:} To run 5 iterations per experiment, you should expect the runtime to take about 7 days when running on an Ubuntu 20.04.5 machine with 19 GiB of RAM and an Intel\textregistered\space Core\texttrademark\space i7-4790 processor running at 3.60 GHz.
              This is primarily due to Wasabi being significantly slower with instrumentation.
        \item \textbf{Publicly available?:} Yes, this artifact is available at the following URL: \url{https://zenodo.org/doi/10.5281/zenodo.10795556}
        \item \textbf{Code license:} Licensed under the Apache License, Version 2.0 (\url{https://www.apache.org/licenses/LICENSE-2.0}).
    \end{itemize}
}

\subsection{Description}

\subsubsection{How to access}
This artifact can be accessed at: \url{https://zenodo.org/doi/10.5281/zenodo.10795556}

\subsubsection{Software dependencies}\label{subsubsec:software-dependencies}
This artifact has the following software dependencies:
\begin{itemize}
    \item \texttt{V8}~\cite{V8}, commit hash \href{https://github.com/v8/v8/commit/f2003219721eb761abe3e5939b254cbf8719291d}{f200321}.
          \texttt{V8} can be downloaded from \url{https://github.com/v8/v8}.
          To run experiments scripts, this binary must be available on the \texttt{PATH} to be called with the \texttt{d8} command.
    \item \texttt{Wasabi}~\cite{Wasabi}, commit hash \href{https://github.com/danleh/wasabi/commit/fe12347f3557ca1db64b33ce9a83026143fa2e3f}{fe12347}.
          \texttt{Wasabi} can be downloaded from \url{https://github.com/danleh/wasabi}.
          To run experiments scripts, this binary must be available on the \texttt{PATH} to be called with the \texttt{wasabi} command.
    \item \texttt{DynamoRIO}~\cite{DynamoRIO}, commit hash \href{https://github.com/pco2699/dynamorio/tree/fc4c25f20db001ec849e10a7ae704d0d108ac684}{fc4c25f}.
          \texttt{DynamoRIO} can be downloaded from \url{https://github.com/DynamoRIO/dynamorio}.
          To run experiments scripts, this binary must be available on the \texttt{PATH} to be called with the \texttt{drrun} command.
    \item Python, version 3.8.10
          To run experiment scripts, both the \texttt{python3} and \texttt{python} (symlinked to \texttt{python3}) commands must be available on the \texttt{PATH}.
    \item Python's venv package, \texttt{python3.8-venv} for \break Debian/Ubuntu systems
    \item \texttt{wasm-bytecode-instrumenter}, commit hash \href{https://github.com/yashanand1910/wasm-bytecode-instrumenter/commit/3ea2003}{3ea2003}.
          The \texttt{bytecode-instrumenter} can be downloaded from \href{https://github.com/yashanand1910/wasm-bytecode-instrumenter}{the repo}.
          To run experiments scripts, this binary must be available on the \texttt{PATH} to be called with the following command: \texttt{wasm-bytecode-instrumenter}
    \item \texttt{rustc}, version 1.71.0.
\end{itemize}

\subsection{Installation and Testing}

\subsubsection{Installation}
To install all required dependencies, follow the detailed instructions in the \texttt{README.md} file in the base directory of the artifact.

\subsubsection{Basic Test}
To verify that an environment is correctly configured to run all scripts provided in this artifact, edit the \texttt{SUITES} variable in the \texttt{common.bash} file to only contain the \texttt{polybench} suite, then run the following command: \texttt{RUNS=2 ./experiment-all-suites.sh}.
We expect this initial test to run in about 1 day (as opposed to the 7 days for all experiments as mentioned above).
When running experiments for \texttt{polybench}, a successful run should result in the CSV directory containing subfolders with CSV output files for each suite script.
Refer to the \texttt{README} for instructions on how to run individual experiments.

\subsection{Experiment workflow}
The workflow of our experiments has two phases: collecting runtime data (saved to \texttt{CSV} files) and generating the corresponding figures (saved to \texttt{PDF} and \texttt{SVG} files).
It is important to remember to save off any data/figures by copying the \texttt{csv}/\texttt{figures} folder to alternate locations prior to running scripts.
If this is not done, the contents will be overwritten.
To collect the runtime data, a user can run any of the \texttt{experiment*.bash} scripts (\texttt{experiment-all-suites.bash} to run all experiments).
Logging information will be output to stdout.
Before plotting data, all experiments should be successfully run (as some figures require data across multiple experiments).
To generate figures, run the \texttt{plot-figure*.py} scripts.

\subsection{Evaluation and expected results}
If you run these experiments, you will find the generated CSV and figure files in their respective folders.
To verify our own results, a side-by-side comparison can be done with the figures in our paper.

\subsection{Experiment customization}
To add your own suite:
\begin{enumerate}
    \item Compile your suite to Wasm.
          The binary can only contain Wasm features supported by the Wasabi and Walrus tools, which tends to be aligned with the core specification.
    \item Make your new suite available in the \texttt{suites} folder following the conventions shown by the other available suites.
    \item Update the \texttt{SUITES} variable in \texttt{common.bash} to contain your new suite.
    \item Update the \texttt{suites} variable in \texttt{plot.py} to contain your new suite.
\end{enumerate}

Helpful variables in \texttt{common.bash}:
\begin{enumerate}
    \item \texttt{RUNS}: configure the number of times run to collect average execution times.
    \item \texttt{SUITES}: configure the suites that will run during experimentation.
\end{enumerate}

\subsection{JVMTI Experiment Artifact}

\subsubsection{Abstract}
We also did a brief experiment discussed in Related Work, Section~\ref{sec:related-work}, to assess the performance overhead imposed by JVMTI's~\cite{JvmTI} handling of \texttt{MethodEntry} events.
To keep our core evaluations separate from this experiment, we have placed this artifact discussion below.
This can be found in the directory \texttt{jvmti} at the base of the artifact located at \url{https://zenodo.org/doi/10.5281/zenodo.10795556}.

\subsubsection{Meta-Information}\label{subsubsec:meta-information}
{\small
\begin{itemize}[leftmargin=2em]
    \item \textbf{Benchmarks:} We leveraged the Richards benchmark for experimenting with JVMTI.
          An equivalent Richards benchmark for Java and Wasm has been provided as part of this artifact.
    \item \textbf{Compilation:} To compile the \texttt{CallsMonitor}, we require \texttt{gcc} version 9.4.0 to be installed.
          To compile and run the Richards Java benchmark, we require Java version 1.8 to be installed.
    \item \textbf{Binaries:} We used the base \texttt{Wizard} binary in our experiment, version \texttt{23a.1617}.
          This binary has been provided as part of the artifact at the location \texttt{bin/base/wizeng.x86-64-linux}.
    \item \textbf{Run-time environment:} We ran this experiment on an \texttt{x86\_64} machine running Ubuntu 20.04.1.
          It is not necessary to have \texttt{sudo} access as software can be installed/symlinked in a user's home directory.
          However, having \texttt{sudo} access would substantially simplify the installation process.
    \item \textbf{Metrics:} Our scripts report \emph{relative execution time} and \emph{absolute overhead} for \texttt{JVMTI} and \oursys.
          To ignore the base startup time required by the engines, we measure instrumented and uninstrumented runs of the Richards benchmark with 0 loops ($Tb_i$ and $Tb_u$ below).
          Given:
          \begin{itemize}
              \item instrumented execution time $T_i$
              \item instrumented base execution time $Tb_i$
              \item uninstrumented execution time $T_u$
              \item uninstrumented base execution time $Tb_u$
          \end{itemize}
          We define \emph{absolute overhead} as the quantity $(T_i-Tb_i)-(T_u-Tb_u)$ and \emph{relative execution time} as the ratio $(T_i-Tb_i)/(T_u-Tb_u)$.
    \item \textbf{Output:} We log all of our output to \texttt{stdout} which should be redirected to a file for inspection.
          To view the summary of each Richard benchmark iteration, \texttt{grep} the file for the term \emph{SUMMARY}.
          The specific iteration of the Richard benchmark is shown in the prefix of each line, e.g. [wasm-9-SUMMARY] means that this line is part of the summary of the Wasm execution of the Richard benchmark with 9 iterations.
          Each iteration is summarized by outputting all execution times for instrumented and uninstrumented variants of the Java and Wasm executions, then reporting the absolute overhead and relative execution time.
          To view the absolute overhead, \texttt{grep} the file for \emph{On average, runtime with monitor took}.
          To view the relative execution time for each Richard benchmark iteration, \texttt{grep} the file for the term \emph{Factor}.
          We have included our own results inside the \texttt{runs\_richards} directory for reference.
    \item \textbf{Experiments:} The artifact scripts run instrumented and uninstrumented variations of the Richards benchmark at 9, 99, 999, 9999, and 99999 loops.
          Each of these variations is run 10x to collect execution time averages across all runs.
    \item \textbf{How much disk space required (approximately)?:} Requires about 2 MB for the \texttt{jvmti} directory and the \oursys engine binary.
    \item \textbf{How much time is needed to prepare workflow (approximately)?:} Shouldn't take longer than 30 minutes since there are few dependencies.
    \item \textbf{How much time is needed to complete experiments (approximately)?:} The experiment takes about 3 hours when running on an Ubuntu 20.04.6 machine with 394 GB of RAM and an Intel\textregistered Xeon\textregistered Platinum 8168 processor running at 2.70 GHz.
    \item \textbf{Publicly available?:} Yes, this artifact is available at the following URL: \url{https://zenodo.org/doi/10.5281/zenodo.10795556}
    \item \textbf{Code license:} Licensed under the Apache License, Version 2.0 (\url{https://www.apache.org/licenses/LICENSE-2.0}).
\end{itemize}}

\subsubsection{Software dependencies}
Running the JVMTI experiment requires following software dependencies:
\begin{itemize}[leftmargin=2em]
    \item Java, version 1.8
    \item \texttt{gcc}, version 9.4.0
\end{itemize}

\subsection{Installation and Testing}

\subsubsection{Installation}
To install all required dependencies to run the scripts, follow the detailed instructions in the \texttt{jvmti/README.md} file.

\subsubsection{Basic Test}
The \texttt{jvmti/README.md} file also describes how to run a basic test to verify your environment setup.

\subsection{Experiment workflow}
The execution workflow is pretty straightforward and outputs all logging information to \texttt{stdout}.
The \texttt{run\_richards.sh} script iterates over the different loops to execute (configured with the \texttt{BENCHMARKS} variable).
For each of these loops to execute, it calculates the average \emph{absolute overhead} and average \emph{relative execution time} over 10 runs (configured with the \texttt{NUM\_RUNS} variable) when running the Richards benchmark for both Java, on the JVM, and Wasm, on Wizard.
If there are issues during execution, errors or warnings will be output.

\subsection{Evaluation and expected results}
Evaluating the results can be done by grep-ing for various types of information as described under ``Output'' in the Meta-Information section above.
The results should be similar to what our own execution found, located in \texttt{jvmti/runs\_richards}.
It is possible that the \emph{absolute overhead} has variations due to differences in the underlying system; however, the \emph{relative execution time} should be similar.

\subsection{Experiment customization}

As described above, it is possible to customize the execution by manipulating the following variables:
\begin{itemize}[leftmargin=2em]
    \item \texttt{NUM\_RUNS}: Fluctuate the number of times each experiment is run to get an average across execution time.
    \item \texttt{BENCHMARKS}: Configure the loops to run on the Richards benchmark.
\end{itemize}

\subsection{Methodology}

Submission, reviewing and badging methodology:

\begin{itemize}[leftmargin=2em]
    \item \url{https://www.acm.org/publications/policies/artifact-review-and-badging-current}
    \item \url{http://cTuning.org/ae/submission-20201122.html}
    \item \url{http://cTuning.org/ae/reviewing-20201122.html}
\end{itemize}

\end{document}